\documentclass[prd, onecolumn, showpacs, floatfix, letterpaper, nofootinbib, amsmath, amssymb, superscriptaddress]
{revtex4}
\usepackage{graphicx}
\usepackage{epsfig}
\usepackage{bm}
\usepackage{amsfonts}
\usepackage{comment}
\usepackage{tikz}
\usetikzlibrary{shapes,arrows}

\usepackage{color}

\newbox\pippobox

\def\be{\begin{equation}}
\def\ee{\end{equation}}
\def\ba{\begin{eqnarray}}
\def\ea{\end{eqnarray}}
\newcommand {\lla} {\ {\raise-.5ex\hbox{$\buildrel<\over\sim$}}\ }

\usepackage[T1]{fontenc}
\usepackage[latin1]{inputenc}
\usepackage{graphicx}
\usepackage[english]{babel}
\usepackage{amsmath}
\usepackage{amssymb}
\usepackage{amsfonts}

\def\be{\begin{equation}}
\def\ee{\end{equation}}
\def\bea{\begin{eqnarray}}
\def\eea{\end{eqnarray}}
\newcommand{\ex}{\mathrm{e}}

\def\spose#1{\hbox to 0pt{#1\hss}}

\def\lta{\mathrel{\spose{\lower 3pt\hbox{$\mathchar"218$}}
     \raise 2.0pt\hbox{$\mathchar"13C$}}}
\def\gta{\mathrel{\spose{\lower 3pt\hbox{$\mathchar"218$}}
     \raise 2.0pt\hbox{$\mathchar"13E$}}}

\newcommand{\Hu}{\mathcal{H}}

\newcommand{\GN}{G_{_\mathrm{N}}}

\begin{document}

\title{Two Field Matter Bounce Cosmology }

\author{Yi-Fu Cai}
\email{yifucai@physics.mcgill.ca}

\author{Evan McDonough}
\email{evanmc@physics.mcgill.ca}

\author{Francis Duplessis}
\email{francis.duplessis@mail.mcgill.ca}

\author{Robert H. Brandenberger}
\email{rhb@hep.physics.mcgill.ca}

\affiliation{Department of Physics, McGill University, Montr\'eal, QC H3A 2T8, Canada}


\begin{abstract}
We re-examine the non-singular Matter Bounce scenario first developed in \cite{BCE}, which starts with a matter-dominated period of contraction and transitions into an Ekpyrotic phase of contraction. We consider both matter fields, the first of which plays the role of regular matter, and the second of which is responsible for the non-singular bounce. Since the dominant matter field is massive, the induced curvature fluctuations are initially not scale-invariant, whereas the fluctuations of the second scalar field (which are initially entropy fluctuations) are scale-invariant. We study the transfer of the initial entropy perturbations into curvature fluctuations in the matter-dominated phase of contraction and show that the latter become nearly scale invariant on large scales but are blue tilted on small scales.  We study the evolution of both curvature and entropy fluctuations through the bounce, and show that both have a scale-invariant spectrum which is blue-tilted on small scales. However, we find that the entropy fluctuations have an amplitude that is much smaller than that of the curvature perturbations, due to gravitational amplification of curvature perturbations during the bounce phase. 
\end{abstract}

\date{\today}
\maketitle

\section{Introduction}

Current high precision data from ground-based \cite{SPT,ACT} and space-based \cite{WMAP, Planck} cosmic microwave background (CMB) telescopes indicate that the origin of structure in the universe is due to a primordial spectrum of nearly adiabatic and nearly scale-invariant cosmological fluctuations. As realized long before these observations \cite{Mukh} (see also \cite{Sato, Press, Starob}), a phase of cosmological inflation during the very early universe will generate such a spectrum. On the other hand, inflation is not the only way to generate such a spectrum. As realized in \cite{Wands, FB}, a scale-invariant spectrum of curvature fluctuations on super-Hubble scales is also generated during a phase of matter-dominated contraction. In order to make contact with the present expanding universe, new physics is required to allow for the transition between the contracting and expanding phases. Such a transition can in principle either be singular and from the point of view of the low-energy effective theory (as in the case of the original Ekpyrotic scenario \cite{Ekp}), or non-singular. There are various ways of obtaining a non-singular bounce, e.g. by modifying the gravitational action as in Horava-Lifshitz gravity \cite{HLbounce}, torsion gravity \cite{Cai:2011tc}, or by adding Null Energy Condition violating matter such as a ghost condensate \cite{ghostcond} or Galileon \cite{Gal} field \footnote{See also \cite{Novello} for a review of bouncing cosmologies.}. A cosmological model with an initial phase of matter-dominated contraction and a non-singular bounce is called the {\it Matter Bounce} scenario and it provides an alternative to cosmological inflation for generating the observed spectrum of cosmological fluctuations (see e.g. \cite{RHBrevs} for review articles on the matter bounce scenario) \footnote{Note that there are other alternatives to inflation for generating a scale-invariant spectrum of cosmological perturbations which, however, will not be discussed in this article.}.

A problem for most bouncing cosmologies is the instability against anisotropic stress, the BKL instability \cite{BKL}. An intuitive way of understanding this problem is to note that the effective energy density in anisotropies evolves with the cosmological scale factor $a(t)$ as $\rho_{anis} \sim a^{-6}$, and thus increases much faster in a contracting universe than the energy densities in matter and radiation. Hence, unless the initial anisotropies are not tuned to zero to a very high precision, no homogeneous bounce will occur.

The solution to this problem, first implemented in the context of the Ekpyrotic scenario \cite{Ekp}, is to introduce a new matter field $\phi$ during the contracting phase whose energy density scales with a higher power of $a^{-1}$ than that of the anisotropy term and which hence dominates the total energy density during the later phases of contraction. With such a field, the BKL instability can be avoided \cite{noBKL}. In \cite{BCE}, a concrete model was proposed in which the new field $\phi$ generates both the Ekpyrotic contraction phase and the non-singular bounce. This is obtained by giving $\phi$ a Galileon-type non-standard kinetic action (which yields the non-singular bounce), and by providing it  with a negative exponential potential which then yields the Ekpyrotic contraction. If we assume that the contracting period starts with a phase of matter-domination, we obtain a realization of the ``matter bounce'' scenario. In \cite{BCE} the evolution of the spectrum of cosmological fluctuations across the bounce phase was studied in detail. In particular, it was shown that the two problems for a certain class of non-singular bounce models discussed in \cite{Stein} do not arise \footnote{The anisotropy remains small during the bounce phase, and there is no dangerous non-scale-invariant fluctuation mode which emerges in the bounce phase.} The stability of this model against anisotropic stress was then confirmed in \cite{Peter} by following the cosmological evolution in the context of an anisotropic Bianchi ansatz.

In the model of \cite{BCE} (and in many other implementations of the ``matter bounce'') there are two matter fields, the field $\phi$ and a field $\psi$ representing the matter which initially dominates the phase of contraction, and which has an equation of state $p = 0$, $p$ denoting the pressure density. Thus, in general there will not only be adiabatic cosmological fluctuations, but also entropic ones. In this paper we give a careful analysis of the evolution of both background and cosmological perturbations in the two field scenario in which a first field $\psi$ generates a matter phase of contraction, and a second field $\phi$ which has a negative exponential potential and hence yields a later phase of Ekpyrotic contraction, and which has a non-trivial kinetic action which generates a non-singular bounce.

We begin in the matter-dominated period of contraction with vacuum fluctuations of both scalar fields. For $\psi$, the resulting power spectrum is blue, since the field has a mass. For $\phi$, the resulting power spectrum on super-Hubble scales is scale-invariant. In the far past, the spectrum of $\phi$ corresponds to the entropy mode, while $\psi$ corresponds to the adiabatic mode. However, at the transition between the matter phase and the Ekpyrotic phase, $\phi$ becomes the adiabatic field, and thus a scale-invariant spectrum of curvature fluctuations results. Due to the gravitational mixing between the two modes during the matter phase of contraction, the $\phi$ fluctuations induce a scale-invariant component to the spectrum of $\psi$ fluctuations at the end of the matter phase of contraction (this is the analog of the ``curvaton'' scenario of structure formation \cite{curvaton} - see also \cite{Yifu}). Hence, the mode which becomes the entropy mode during the later phases of evolution also inherits a scale-invariant contribution in addition to the original contribution which has a steep blue spectrum.

The outline of the paper is as follows: in Section II we discuss the model for a non-singular bounce proposed in \cite{BCE}, and how this is affected by the addition of an additional scalar field of K-essence form. In Section III we describe the background cosmological evolution by splitting the time history of the universe into phases: matter contraction, Ekpyrotic contraction, non-singular bounce, and fast roll expansion. To justify this phase structure, and to serve as a evidence that this model is feasible, we study the background numerically. In Section IV, we consider the evolution of perturbations our model, which we then use in Section V to calculate the power spectra at late
times. We finish with some concluding remarks in Section VI.

A word on notation: We define the reduced Planck mass by $M_p  = 1/\sqrt{8\pi \GN}$ where $\GN$ is Newton's gravitational constant. The sign of the metric is taken to be $(+,-,-,-)$. Note that we take the value of the scale factor at the bounce point to be $a_{B} = 1$ throughout the paper.

\section{Cosmology of a Non-Singular Bounce}

As discussed in the introduction, the model of interest for the present work is that of two scalar fields: a matter field which dominates at very early times, and a bounce field which violates the Null Energy Condition for a brief period, inducing the bounce.  We begin with the most general Lagrangian for this class of models, given by
\begin{eqnarray}\label{Lagrangian}
 \mathcal{L} = K(\phi, X) + G(\phi, X) \Box\phi + P(\psi, Y) ~,
\end{eqnarray}
where $\phi$ is the bounce field of Galilean type, $\psi$ is a K-essence scalar of general form, and have defined
\begin{eqnarray}
 X \equiv \frac{1}{2}\partial_\mu\phi\partial^\mu\phi~,~~
 Y \equiv \frac{1}{2}\partial_\mu\psi\partial^\mu\psi~ ,
\end{eqnarray}
as well as the d'Alembertian operator
\begin{equation}
 \Box \equiv g^{\mu\nu} \nabla_\mu \nabla_\nu  .
\end{equation}
The Lagrangian terms for the bounce field are defined as
\begin{eqnarray}\label{Kessence}
 K(\phi, X) = M_p^2 \left[1-g(\phi) \right]X + \beta {X}^2 - V(\phi),
\end{eqnarray}
\begin{eqnarray}\label{Galileon}
 G(\phi,X) = \gamma X,
\end{eqnarray}
where we have parametrized the model via the positive-definite constants\footnote{The  positive-definiteness of $\beta$ ensures that the kinetic term is bounded from below at high energy scales} $\beta$ and $\gamma$, as well as the functions $g(\phi)$ and $V(\phi)$. The term $G(\phi, X)$ is a Galileon-type operator which we have introduced to stabilize the gradient term of cosmological perturbations,  and leads to a sound speed which is positive-definite at all time except for during the bounce. Note that we have adopted the convention that $\phi$ is dimensionless, and so we include a factor of $M^{2} _{pl}$ in $K(\phi,X)$.

The bounce is triggered when $g(\phi)<1$, which causes $\phi$ to form a ghost condensate and hence violate the Null Energy Condition. The function is negligible far from the bounce, such that the bounce field $\phi$ will have canonical kinetic terms at early and late times, given suitable behaviour for $X$. We can build this function by setting the bounce to occur at $\phi=0$, and requiring that $g<1$ when $|\phi| \gg 1$ but $g>1$ when $\phi \sim 0$. We choose its form to be
\begin{eqnarray}\label{gphi}
 g(\phi) = \frac{2g_0}{\ex^{-\sqrt{\frac{2}{p}}\phi}+\ex^{b_g\sqrt{\frac{2}{p}}\phi}},
\end{eqnarray}
where $g_0 \equiv g(0)$ and $p$ are positive constants, with $g_0$ larger than unity, $g_0>1$ and $p$ smaller than unity, $p<1$. 

The bounce field potential $V(\phi)$  is chosen to ensure that the bounce is preceded by a phase of Ekpyrotic contraction, which is necessary to dilute anisotropy and avoid the BKL instability. 
The potential can also be chosen to give an attractor solution in both the expanding and contracting branches of the cosmological evolution, by making use of exponential functions. We take the form of the potential to be
\begin{eqnarray}\label{Vphi}
 V(\phi) = -\frac{2V_0}{\ex^{-\sqrt{\frac{2}{q}}\phi}+\ex^{b_V\sqrt{\frac{2}{q}}\phi}},
\end{eqnarray}
where $V_0$ is a positive constant with dimension of $({\rm mass})^4$,   $q$ is a positive constant that must be smaller than $1/3$ in order to obtain Ekpyrotic contraction, and the constant $b_V$ is an asymmetry parameter for the potential. The attractor solution is induced during expansion by the positive-valued exponential, while the negative exponential leads to an attractor solution in the contracting phase.

We now turn to the second field $\psi$, which we introduce to play the role of an arbitrary matter field satisfying the Null Energy Condition. Initially, we take its Lagrangian to be of K-essence form, $P(\psi, Y)$, but eventually we will consider a canonical massive free scalar field. It has pressure and energy density given by
\begin{eqnarray}
 p_\psi &=& P~,\\
 \rho_\psi &=& 2YP_{,Y}-P~.
\end{eqnarray}
There are two important quantities for this system: the equation of state $w_{\psi}$ and the sound speed square $c_\psi^2$. These are given by
\begin{eqnarray}
\label{eos_K}
 w_\psi &\equiv& \frac{p_\psi}{\rho_\psi} = -1 + \frac{2YP_{,Y}}{2YP_{,Y}-P}~,\\
\label{cs2_K}
 {c_\psi^2} &\equiv& \frac{{p_\psi}_{,Y}}{{\rho_\psi}_{,Y}} = \frac{P_{,Y}}{2YP_{,YY}+P_{,Y}}~.
\end{eqnarray}

We now consider the spatially flat FRW universe whose metric is given by
\begin{eqnarray}\label{FRWmetric}
 ds^2 \, = \, dt^2-a^2(t)d\vec{x}^2~,
\end{eqnarray}
where $t$ is cosmic time, $x$ are the comoving spatial coordinates and $a(t)$ is the scale factor. The evolution of the scale factor can be characterized by the Hubble rate:
\begin{eqnarray}
 H \equiv \frac{\dot{a}}{a}~,
\end{eqnarray}
where the dot denotes the derivative with respect to cosmic time $t$.

At the background level the universe is homogenous, and thus both the bounce field $\phi$ and the matter field $\psi$ are only functions of cosmic time. Thus, the kinetic terms of these two fields become
\begin{eqnarray}
 X = \dot\phi^2/2 ~,~~
 \Box\phi = \ddot\phi + 3H\dot\phi ~,~~
 Y = \dot\psi^2/2~.
\end{eqnarray}
The pressure and energy density of the bounce field are given by
\begin{eqnarray}
\label{pressure}
 p_\phi &=& \frac{1}{2}M_p^2 (1-g)\dot\phi^2 +\frac{1}{4}\beta\dot\phi^4 -\gamma\dot\phi^2\ddot\phi -V(\phi)~,\\
\label{rho}
 \rho_\phi &=& \frac{1}{2}M_p^2 (1-g)\dot\phi^2 +\frac{3}{4}\beta\dot\phi^4 +3\gamma H\dot\phi^3 +V(\phi)~,
\end{eqnarray}
where dynamics of $\phi$ are governed by the equation of motion
\begin{eqnarray}\label{eom}
 {\cal P} \ddot\phi + {\cal D} \dot\phi +V_{,\phi} = 0~,
\end{eqnarray}
and we have introduced
\begin{eqnarray}
\label{Pterm}
 {\cal P} &=& (1-g) M_p^2 +6\gamma H\dot\phi +3\beta\dot\phi^2 +\frac{3\gamma^2}{2M_p^2}\dot\phi^4,\\
\label{Fterm}
 {\cal D} &=& 3(1-g) M_p^2 H +\left(
 9\gamma{H}^2-\frac{1}{2} M_p^2 g_{,\phi}\right)
 \dot\phi +3\beta{H}\dot\phi^2 \nonumber\\
 && -\frac{3}{2}(1-g)\gamma\dot\phi^3 -\frac{9\gamma^2H\dot\phi^4}{2M_p^2} -\frac{3\beta\gamma\dot\phi^5}{2M_p^2}  - \frac{3G_{,X}}{2M_p^2}(\rho_\psi+p_\psi)\dot\phi ~.
\end{eqnarray}
From Eq.~\eqref{eom}, it is clear that the function ${\cal P}$ determines the positivity of the kinetic term of the scalar field and thus can be used to determine whether the model contains a ghost or not at the perturbative level; the function ${\cal D}$ on the other hand, represents an effective damping term. By keeping the first terms of the expressions for ${\cal P}$ and ${\cal D}$ and setting $g = 0$, which is a good approximation far from the bounce where $\dot{\phi} \ll M_{Pl}$, one can recover the standard Klein-Gordon equation in the FRW background. Note that the friction term ${\cal D}$ contains the contributions from the matter fluid, which can be suppressed for small values of $\dot\phi$. However, these terms will become important during the bounce phase where $\dot\phi$ reaches a maximal value.

For completeness, we can write down the Einstein equations in this background,
\begin{eqnarray}
 M_p^2 \left(R_{\mu\nu}-\frac{R}{2}g_{\mu\nu} \right) = T^\phi_{\mu\nu} +T^\psi_{\mu\nu} ,
\end{eqnarray}
and the corresponding Friedmann equations,
\begin{eqnarray}
\label{Friedmann1} H^2 &=& \frac{\rho_{_\mathrm{T}}}{3M_p ^2} ~,\\
\label{Friedmann2} \dot{H} &=& -\frac{\rho_{_\mathrm{T}} + p_{_\mathrm{T}}}{2M_p ^2} ~,
\end{eqnarray}
where $\rho_{_\mathrm{T}}$ and $p_{_\mathrm{T}}$ represent the total energy density and pressure in the FRW universe, e.g. the sum of the contributions of the bounce field and the matter field.

\section{Background evolution}

The initial conditions of the background are chosen such that the universe is initially dominated by regular matter in the contracting phase, which in our model is mimicked by the matter field $\psi$. Since the potential of the bounce field $V(\phi)$ has an Ekpyrotic potential for $\phi\ll1$, the corresponding energy density grows faster than that of regular matter. As a consequence, $\phi$ eventually becomes dominant, signaling the end of matter contraction. After that, the Ekpyrotic phase of contraction begins, and lasts until the non-singular bounce interval begins (this is the phase where the effects coming from new physics dominate), followed by a period of fast-roll expansion, which in turn ends at a transition to the expansion of Standard Big Bang cosmology. We choose the initial conditions for the density of regular matter and for the value of $\phi$ such that the temperature at which the Ekpyrotic phase begins is higher than that at the time of equal matter and radiation in the Standard Big Bang expanding phase.

\subsection{Analytic estimates}

In the following we briefly investigate the evolution of the universe in each of the periods mentioned above, and refer to \cite{Peter} for a more generic analysis in which the anisotropy was taken into account as well.

\subsubsection{Matter contraction}

We start by considering the period when the universe is dominated by the matter field $\psi$. We take the Lagrangian of $\psi$ to be that of a free canonically normalized massive scalar field:
\begin{eqnarray}
 P(\psi, Y) = Y-\frac{1}{2}m^2\psi^2~.
\end{eqnarray}
Thus the matter field oscillates around its vacuum state $\psi=0$ and the time-averaged background equation of state parameter is roughly $w=0$. In this phase, the scale factor evolves as
\begin{eqnarray}
 a(t) \simeq a_{E} \left( \frac{t-\tilde{t}_{E}}{t_{E} -\tilde{t}_{E}} \right)^{2/3},
\end{eqnarray}
where $t_{E}$ denotes the final moment of matter contraction and the beginning of the Ekpyrotic phase, and $a_{E}$ is the value of the scale factor at the time $t_{E}$. In the above, $\tilde{t}_{E}$ is an integration constant which is introduced to match the Hubble parameter continuously at the time $t_{E}$,
\begin{eqnarray}
 \tilde{t}_{E} \simeq t_{E}-\frac{2}{3H_{E}}.
\end{eqnarray}
Hence the Hubble parameter can be approximated by
\begin{eqnarray}\label{Hubble_c}
 \langle H(t) \rangle = \frac{2}{3(t-\tilde{t}_{E})}~.
\end{eqnarray}
where the angular brackets stand for averaging over time. The solution for the scalar field $\psi$ can be asymptotically expressed (modulo a phase) as
\begin{eqnarray}
 \psi(t) \simeq \tilde{\psi}(t) \sin(m(t-\tilde{t}_{E}))~,
\end{eqnarray}
with a time dependent amplitude
\begin{eqnarray}
 \tilde{\psi}(t) = \frac{1}{\sqrt{3\pi G}m(t-\tilde{t}_{E})} ~,
\end{eqnarray}
which yields an equation of state with has vanishing pressure when averaged over an oscillation period of the field.

\subsubsection{Ekpyrotic contraction}

We assume a homogeneous scalar field $\phi$ which is initially placed in the region $\phi \ll -1$ in the phase of matter contraction. In this case, the Lagrangian for $\phi$ approaches the conventional canonical form. Once $\phi$ begins to dominate the energy-momentum tensor of matter, it then approaches an attractor solution which is given by
\begin{eqnarray}
 \phi(t) \simeq -\sqrt{\frac{q}{2}} \ln \left[ \frac{2V_0(t-\tilde{t}_{B-})^2}{q(1-3q)M_p ^2} \right] ~,
\end{eqnarray}
where $\tilde{t}_{B-}$ is an integration constant which chosen such that the Hubble parameter at the end of the phase of Ekpyrotic contraction matches with the one at the beginning of the bounce phase. This attractor solution corresponds to an effective equation of state
\begin{eqnarray}\label{eos_ekpy}
 w \simeq -1+\frac{2}{3q} ~.
\end{eqnarray}

During the phase of Ekpyrotic contraction, the scale factor evolves as
\begin{eqnarray}
 a(t) \simeq a_{B-} \left(\frac{t-\tilde{t}_{B-}}{t_{B-}
 -\tilde{t}_{B-}}\right)^q ~,
\end{eqnarray}
where $a_{B-}$ is the value of scale factor at the time $t_{B-}$ which corresponds to the end of Ekpyrotic contraction and the beginning of the bounce phase. Therefore, the Hubble parameter is given by
\begin{eqnarray}\label{Hubble_E}
 H(t) \simeq \frac{q}{t-\tilde{t}_{B-}} ~,
\end{eqnarray}
where, in order to make $H(t)$ continuous at the time $t_{B-}$, one must set
\begin{eqnarray}
 \tilde{t}_{B-} = t_{B-}-\frac{q}{H_{B-}} ~.
\end{eqnarray}
Additionally, we require the scale factor to evolve smoothly and continuously at the time $t_{E}$. This leads to the relation
\begin{eqnarray}\label{a_E}
 a_{E} \simeq a_{B-} \left(\frac{H_{B-}}{H_{E}}\right)^q ~.
\end{eqnarray}

\subsubsection{Bounce phase}

In our model the scalar field evolves monotonically from $\phi \ll -1$ to $\phi \gg 1$. For values of $\phi$ between $\phi_- \sim -\sqrt{p/2} \ln (2g_0)$ and $\phi_+\sim \sqrt{p/2} \ln (2g_0)/b_g$ (assuming one term in the denominator of $g(\phi)$ dominates over the other at each transition time), the value of the function $g(\phi)$ becomes larger than unity and thus the universe enters a ghost condensate state. The occurrence of the ghost condensate naturally yields a short period of Null Energy Condition violation and this in turn gives rise to a non-singular bounce \cite{ghostcond}.

As shown in Ref. \cite{BCE}, we have two useful parameterizations to describe the evolution of the scale factor in the bounce phase. One is the linear parametrization of the Hubble parameter
\begin{eqnarray}\label{Hubble_bounce}
 H(t) \simeq \Upsilon t ~,
\end{eqnarray}
and the other is the evolution of the background scalar
\begin{eqnarray}\label{dotphi_bouncing}
 \dot\phi(t) \simeq \dot\phi_{B} \ex^{-t^2/T^2} ~,
\end{eqnarray}
where the coefficient $\Upsilon$ is set by the detailed microphysics of the bounce. The coefficient $T$ can be determined by matching the detailed evolution of the scalar field at the beginning or the end of the bounce phase, which will be addressed in next subsection. Thus, during the bounce the scale factor evolves as
\begin{eqnarray}
 a(t) \simeq a_{B} \ex^{\frac12 \Upsilon t^2} ~.
\end{eqnarray}

Note that a non-singular bounce requires that the total energy density vanishes at the bounce point. The total energy density includes the contributions from the matter fields and the anisotropy factors. This leads to the following result for the value of $\dot\phi_{B}$
\begin{eqnarray}\label{dot_phi_B^2}
 \dot\phi_{B}^2 &\simeq& \frac{(g_0-1)M_p^2}{3\beta} \left[ 1+\sqrt{1+\frac{12\beta(V_0+\rho_\mathrm{m}+\rho_\theta)}{(g_0-1)^2 M_p^4}} \,\right] \nonumber\\
 &\simeq& \frac{2(g_0-1)}{3\beta} M_p^2 ~,
\end{eqnarray}
where we have made use of approximations that $\rho_\mathrm{m}$ and $\rho_\theta$ are much less than $V_0$ and $V_0\ll M_p^4$ in the second line. These approximations must be valid for the model to hold since both $\rho_\mathrm{m}$ and $\rho_\theta$ are greatly diluted in the Ekpyrotic phase and $V_0$ is the maximal absolute value of the potential of $\phi$ which, according to the observational constraint from the amplitude of cosmological perturbations, must be far below the Planck scale.

\subsubsection{Fast-roll expansion}

After the bounce, the universe enters the expanding phase, where the universe is still dominated by the scalar field $\phi$. During this stage, the motion of $\phi$ is dominated by its kinetic term while the potential is negligible. Thus, the background equation of state parameter is $w \simeq 1$. This corresponds to a period of fast-roll expansion, where the scale factor evolves as
\begin{eqnarray}
 a(t) \simeq a_{B+} \left(\frac{t-\tilde{t}_{B+}}
 {t_{B+}-\tilde{t}_{B+}}\right)^{1/3} ~,
\end{eqnarray}
where $t_{B+}$ represents the end of the bounce phase and the beginning of the fast-roll period, and $a_{B+}$ is the value of the scale factor at that moment. Then one can write down the Hubble parameter in the fast-roll phase
\begin{eqnarray}
 H(t) \simeq \frac{1}{3(t-\tilde{t}_{B+})} ~,
\end{eqnarray}
and the continuity of the Hubble parameter at $t_{B+}$ yields
\begin{eqnarray}
 \tilde{t}_{B+} = t_{B+} -\frac{1}{3H_{B+}} ~.
\end{eqnarray}

Recall that, in Eq. \eqref{dotphi_bouncing}, we made use of a Gaussian parametrization of the scalar field evolution in the bounce phase, with characteristic timescale $T$. In the fast roll phase we find the following approximate solution for the evolution of $\phi$:
\begin{eqnarray}
 \dot\phi(t) \simeq \dot\phi_{B+} \frac{a_{B+}^3}{a^3(t)}
 \simeq \dot\phi_{B} \ex^{-{t_{B+}^2}/{T^2}}\frac{H(t)}{H_{B+}} ~,
\end{eqnarray}
where we have applied \eqref{dotphi_bouncing} in the second equality. This implies that
\begin{eqnarray}
 \rho_\phi \simeq \frac{M_p^2}{2}\dot\phi^2 \simeq \frac{M_p^2\dot\phi_{B}^2}
 {2\ex^{2t_{B+}^2/T^2}} \frac{H^2}{H_{B+}^2} ~.
\end{eqnarray}
Moreover, the Friedmann equation requires that $\rho_\phi\simeq 3M_p^2H^2$ in the fast-roll phase, so that $T^2$ is given by
\begin{eqnarray}
 T^2 \simeq \frac{2H_{B+}^2}{\Upsilon^2\ln\left[\displaystyle
 \frac{M_p^2(g_0-1)}{9\beta H_{B+}^2}\right]} ~.
\end{eqnarray}

\subsection{A (Numerical) Proof of Principle}

To justify our claims that the background does exhibit this phase structure, we numerically solve the background equations of motion. We present this solely as a `Proof of Principle', in order to illustrate the occurrence of a non-singular bounce in the model under consideration. By this we mean that the parameters are chosen to make the effect of the matter field $\psi$ manifest during the bounce, but this parameter choice does not necessarily satisfy the bounds imposed by observations. Assuming parameter values taking into account the experimental constraints would lead to an Ekpyrotic phase which is long enough to dilute all the matter fields, which would decrease the significance of entropy perturbations. Similarly, in the limit that the Ekpyrotic phase stretches to the infinite past, the evolution of the background approaches that obtained in a regular isotropic bounce model realized by a single field as studied in \cite{BCE}.

In the numerical calculation we work in units of the Planck mass $M_p$ for all variables. We specifically set a group of model parameters as,
\begin{eqnarray}\label{parameters1}
 & V_0=10^{-10}~,~~ g_0=1.1~,~~ \beta=5~,~~ \gamma=10^{-3}~,~~ \nonumber\\
 & b_V=5~,~~ b_g=0.5~,~~ p=0.01~,~~ q=0.1~,~~ m = 5\times 10^{-6}~.
\end{eqnarray}
Moreover we choose the initial conditions for the bounce field and matter field as follows,
\begin{eqnarray}\label{parameters2}
 & \phi_{\rm ini} = -2.11~,~~ \dot\phi_{\rm ini}= -8.87\times 10^{-8}~, \nonumber\\
 & \psi_{\rm ini} = -0.025~,~~ \dot\psi_{\rm ini}= -3.57\times 10^{-8}~.
\end{eqnarray}

\begin{figure}
\includegraphics[scale=0.5]{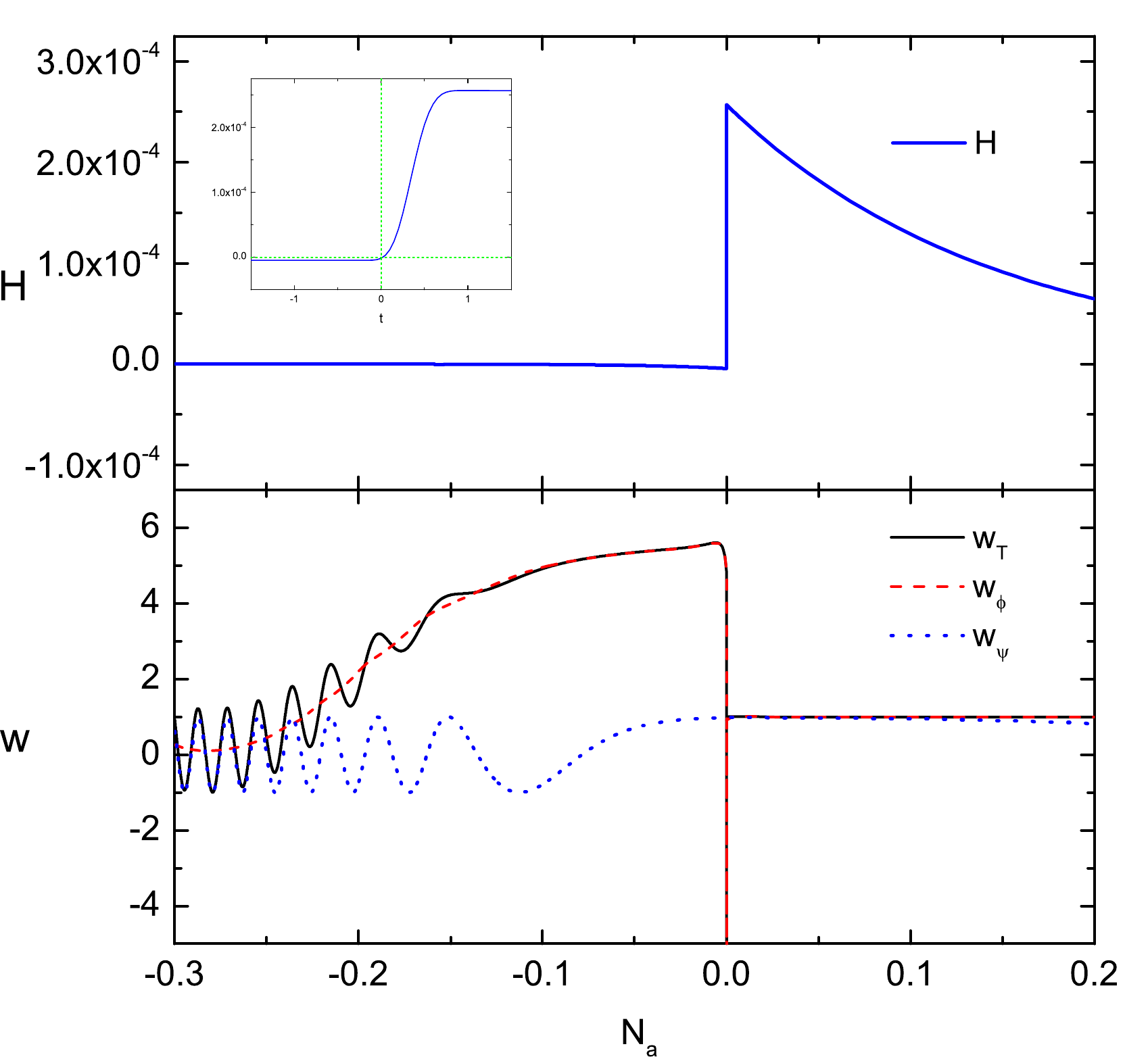}
\caption{Cosmic evolution of the Hubble parameter $H$ (blue line in the upper panel) and the equations of state (black solid, red dashed, and blue dotted lines in lower panel for the total background $w_{_\mathrm{T}}$, the bounce field $w_\phi$ and the matter field $w_\psi$, respectively), in units of the reduced Planck mass $M_p$, with background parameters given by \eqref{parameters1} and initial conditions as in \eqref{parameters2}. The main plot shows that a non-singular bounce occurs, and that the time scale of the bounce is short (it is a ``fast bounce'' model). The inner insert shows a zoomed-in view of the smooth Hubble parameter during the bounce phase as a function of cosmic time. }
\label{Fig:HEoS}
\end{figure}

\begin{figure}
\includegraphics[scale=0.5]{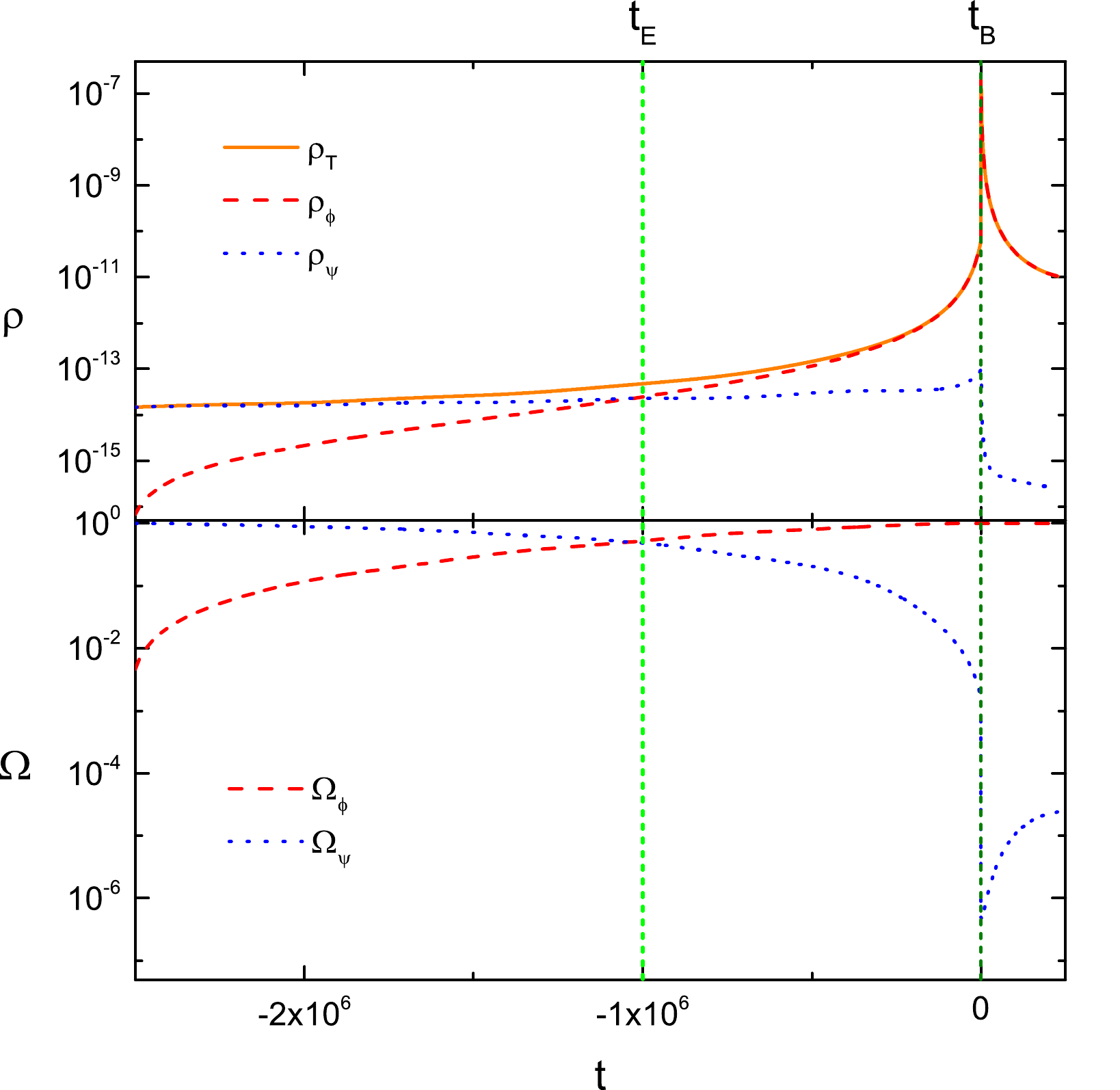}
\caption{Cosmic evolution of the energy densities $\rho$ and density parameters $\Omega$ of the background universe (orange solid line in upper panel), the bounce field (red dashed line) and the matter field (blue dotted line), respectively. The horizontal axis is the cosmic time. The initial conditions and model parameters are the same as in Fig. \ref{Fig:HEoS}. }
\label{Fig:rhoOmega}
\end{figure}

Our numerical results are presented in Figs. \ref{Fig:HEoS} and \ref{Fig:rhoOmega}. In order to enlarge the details of the cosmic evolution, we introduced a parameter
\begin{eqnarray}
 N_a \equiv \left\{ \begin{array}{c}
            -\ln ~\frac{a}{a_0} ~~t<t_B \\
            \\
            \ln ~\frac{a}{a_0} ~~t\geq t_B
\end{array} \right.
\end{eqnarray}
(where $a_0$ is a normalization constant) as the horizontal axis in Fig. \ref{Fig:HEoS}. The vertical axis shows the dynamics of the Hubble parameter and the equations of state of scalar fields as well as the overall one.

From the upper panel of Fig. \ref{Fig:HEoS}, one can see that the Hubble parameter evolves smoothly through the bounce point with an approximately linear dependence on cosmic time. However, the bounce phase is not symmetric with respect to the bounce point in this model. The lower panel of Fig. \ref{Fig:HEoS} shows that the background equation of state initially takes an average value $w=0$ since the universe is dominated by the oscillating matter field $\psi$. During the matter contraction, the bounce field slowly becomes dominant over and triggers a period of Ekpyrotic contraction, where for our parametrization the equation of state is approximately equal to $w=5.67$. When the universe enters the bounce phase, the background equation of state experiences a sudden decrease to negative infinity and then evolves back to a value $w=1$ which signals a fast-roll expanding phase.

In order to better characterize the transitions between different phases, we plot the evolution of the energy densities and density parameters in Fig. \ref{Fig:rhoOmega}. The density parameters are defined as
\begin{eqnarray}
 \Omega_i \equiv \frac{\rho_i}{\rho_{_\mathrm{T}}}~,
\end{eqnarray}
where the subscript ``$i$" represents $\phi$ and $\psi$, respectively. This figure explicitly shows that the universe in this model experiences four phases: Matter contraction, Ekpyrotic contraction, the bounce, and fast-roll expansion.

\section{Cosmological Perturbations}

\subsection{Overview}

In this section we study the dynamics of linear cosmological perturbations in the Two Field Matter Bounce. One attractive property of a non-singular bounce cosmology is that perturbation modes can be evolved smoothly through the bounce phase. In linear theory, perturbations of scalar type evolve independently from those of vector and tensor type. This reduces the number of degrees of freedom which must be analyzed. In addition, as a consequence of linearity one can track each Fourier mode independently (see e.g. \cite{MFB} for a survey of the theory of cosmological perturbations and \cite{RHBpertrev} for an introductory overview). The evolution of the Fourier modes depends on the background cosmology.

As per the analysis presented in the previous section, our cosmological background will first undergo matter contraction, then a period of Ekpyrotic contraction, followed by a non-singular bounce, and then a phase of fast roll expansion. We begin with vacuum fluctuations on sub-Hubble scales in contracting phase. During the phase of contraction, wavelengths exit the Hubble radius (which is shrinking in comoving coordinates). Once they are on super-Hubble scales, the modes are squeezed. Both the exiting of the Hubble radius and the squeezing on super-Hubble scales is similar to what happens during the phase of accelerated expansion in inflationary cosmology. However, in the case of inflation the Hubble length has constant physical size while the physical wavelength of fluctuations increases exponentially. Hence, if the period of inflation was long, the physical wavelength of the fluctuations was initially smaller than the Planck length, leading to the `trans-Planckian problem' for fluctuations \cite{Jerome}. This problem does not arise in a bouncing cosmology as long as the energy scale of the bounce is smaller than the Planck scale, as is required for the self-consistency of any effective field treatment such as what we are presenting, since then the physical wavelength of the fluctuation modes which we measure today were always much larger than the Planck length scale.

As was initially realized in \cite{Wands, FB},  a consequence of the super-Hubble growth of fluctuations in a contracting universe is that the initial vacuum fluctuations of a massless scalar field (and consequently also the curvature fluctuations in a model in which the only matter component is this massless field) are converted to a scale-invariant spectrum. It is in this sense that the matter bounce can provide an alternative to inflationary cosmology as a mechanism to form the cosmological fluctuations we observe today.

However, the model under consideration involves two scalar fields, $\phi$ and $\psi$, with $\psi$ leading to a phase of a matter contraction at early times, and $\phi$ being responsible for the Ekpyrotic phase and the bounce. The dominant field in the initial matter phase of contraction is massive and hence its vacuum spectrum does not evolve into a scale-invariant form in isolation. The field $\phi$, on the other hand, is effectively massless at early times and hence evolves to a scale-invariant spectrum on super-Hubble scales. The field $\phi$ acts an entropy field during the phase of matter contraction. However, once the Ekpyrotic phase begins, $\phi$ becomes dominant and becomes the curvature mode, while the $\psi$ fluctuations become the entropy modes.

As is well known, entropy modes source a growing curvature perturbation on super-Hubble scales \footnote{See e.g. \cite{Minos} for an early discussion in the context of an axion dominated inflationary universe.} Thus, to determine the final spectrum of curvature and entropy fluctuations in our model we must carefully study the interaction of the two fluctuation modes in each cosmological phase. As we will show, in the matter phase of contraction, the scale-invariant $\phi$ mode (which acts as an entropy fluctuation) seeds a curvature fluctuation (the $\psi$ mode in the initial phase) of comparable magnitude. Thus, at the end of the matter-dominated phase of contraction, both modes are scale-invariant and have comparable amplitude. After that time, it is no longer important to consider the sourcing of the adiabatic mode by the entropy mode since the adiabatic mode is already larger in amplitude (and the effect of the sourcing cannot induce a larger amplitude than that of the source)

In many non-singular bounce models it has been shown that the scale-invariance of curvature fluctuations is preserved during the bounce phase (see, however, the exceptions discussed in \cite{Stein}). We will show that this is also the case in our model. We will also evolve the entropy fluctuations on super-Hubble scales and will show that they preserve their scale-invariance on large scales. Moreover the curvature mode are amplified compared to the the entropy mode during the bounce phase, and thus the final spectrum of fluctuations is almost completely adiabatic.

In both the matter contraction phase and the Ekpyrotic phase, the Lagrangian of the bounce scalar recovers the canonical form, since the higher derivative terms are suppressed by the small value of $\dot\phi$. In the matter contraction phase, it is convenient to study the evolution of perturbation modes in the spatially flat gauge ($\zeta=0$) and the initial conditions for two field fluctuations can be imposed inside the Hubble radius. However once the initial conditions have been set, we can switch into the uniform $\phi$ gauge ($\delta\phi=0$) for the Ekpyrotic and subsequent phases. In this way,  the curvature perturbation becomes manifest.

To perform this perturbation analysis we use three sets of perturbation variables. For the initial conditions, we consider the field fluctuations in the spatially flat gauge

\begin{equation}
\label{MSvars}
Q_\phi = M_p(\delta \phi + \frac{\dot{\phi}}{H} \Phi)~ \;\;\;\;\; , \;\;\;\;\; Q_\psi =   \delta \psi + \frac{\dot{\psi}}{H} \Phi~ .
\end{equation}
where $\Phi$ is the Bardeen potential (see Appendix A). We can change to the uniform $\phi$ gauge, where the perturbation variables become $\delta \psi$ and
\begin{equation}
\label{zeta}
\zeta =  H \frac{( M_p\dot{\phi} Q_\phi + \dot{\psi} Q_\psi )}{M_p^2\dot{\phi}^2 +\dot{\psi}^2}~ \;\;\;\;\;\;\;\;\;\; \delta \psi \rightarrow Q_\psi \, .
\end{equation}

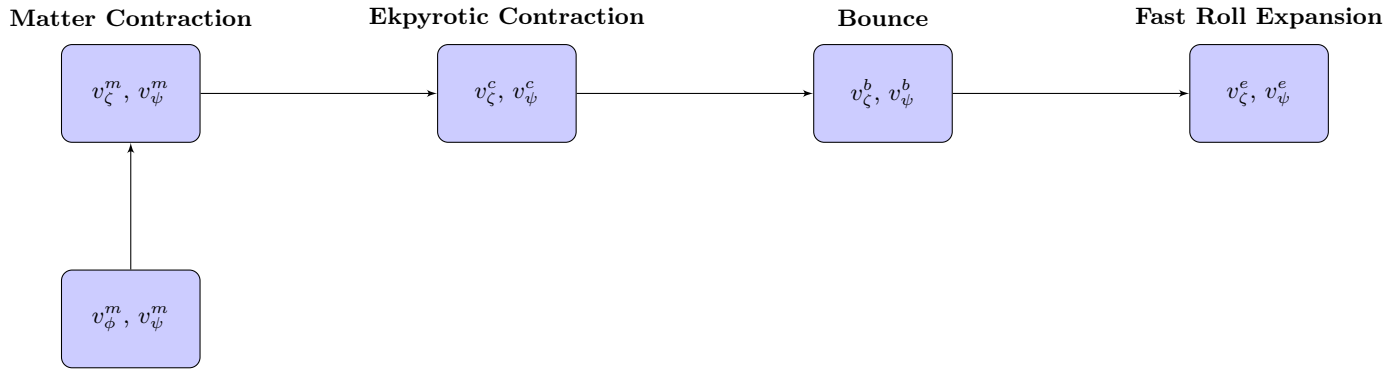
\begin{figure}

\label{variables}
\caption{Stages of the Perturbation analysis. We begin in the matter dominated phase by using the field fluctuations, then change gauge to study the entropy and curvature perturbations. We match the curvature perturbations through the bounce to solve for the behaviour in the fast roll expansion phase. We denote the phases of the bounce by indices on the perturbation variables: $m$ , $c$ , $b$, and $e$, for {\bf m}atter domination, Ekpyrotic {\bf c}ontraction, {\bf b}ounce, and fast roll {\bf e}xpansion.}

\tikzstyle{decision} = [diamond, draw, fill=blue!20,
    text width=4.5em, text badly centered, node distance=3cm, inner sep=0pt]
\tikzstyle{block} = [rectangle, draw, fill=blue!20,
    text width=5em, text centered, rounded corners, minimum height=4em]
\tikzstyle{line} = [draw, -latex']
    \begin{centering}
\begin{tikzpicture}[node distance = 2cm, auto]
    \node [block] (1) { $v^m_{\zeta}$, $v^m_{\psi}$};
    \node [above of =1, node distance = 1cm] {{\bf Matter  Contraction}};
    \node [block, below of=1, node distance=3cm] (2) { $v^m_{\phi}$, $v^m_\psi$};
    \node [block, right of=1, node distance=5cm] (3) { $v^c_{\zeta}$, $v^c_{\psi}$};
    \node [above of =3, node distance = 1cm] {{\bf  Ekpyrotic Contraction}};
    \node [block, right of=3, node distance=5cm] (4) { $v^b_{\zeta}$, $v^b_{\psi}$};
    \node [above of =4, node distance = 1cm] {{\bf Bounce}};
    \node [block, right of=4, node distance=5cm] (5) { $v^e_{\zeta}$, $v^e_{\psi}$};
    \node [above of =5, node distance = 1cm] {{\bf Fast Roll Expansion}};

    \path [line] (2) -- (1);
    \path [line] (1) -- (3);
    \path [line] (3) -- (4);
    \path [line] (4) -- (5);

\end{tikzpicture}

\end{centering}
\end{figure}

We lay out the general recipe for the perturbation analysis in Figure \ref{variables}

\subsection{Field fluctuations during matter contraction}

At the beginning of matter contraction, the universe is dominated by the matter field $\psi$ which is oscillating around its vacuum point; this yields a time averaged value of the background equation of state $w\simeq 0$ and thus the universe is in a matter dominated phase. During this phase, the bounce field $\phi$ is subdominant and fast rolling down along its potential with an effective equation of state $w_\phi\simeq 1$.

One can perturb the metric and the two scalar field to linear order, which includes three scalar type perturbation modes, $\zeta$, $\delta\phi$ and $\delta\psi$, respectively. However, one of these three variable can be eliminated by making a gauge choice. We start by considering the evolution of cosmological perturbations using the gauge invariant field fluctuations $Q_\phi$ and $Q_\psi$ defined in Eq. (\ref{MSvars}), which are the Mukhanov-Sasaki variables\cite{Sasaki:1986hm, Mukhanov:1988jd}.

One can introduce the gauge invariant curvature perturbation as in Eq. (\ref{zeta}), as well as the entropy perturbation
\begin{equation}
 {\cal S} = \frac{ \left(M_p\dot{\phi} Q_\psi - \dot{\psi} Q_\phi \right)}{\sqrt{M_p^2\dot{\phi}^2 +\dot{\psi}^2}}~.
\end{equation}

At early times in the matter dominated phase, $\dot{\psi} \gg \dot{\phi}M_p$, which implies that
\begin{eqnarray}
 \zeta \simeq H\frac{Q_\psi}{\dot\psi}~,~~ {\cal S} \simeq - Q_{\phi}~  \;\;\;\; \mbox{for}\;\; t \rightarrow -\infty .
\end{eqnarray}
Therefore, one can immediately observe that at very early times the main contribution to the curvature perturbation is from the matter field fluctuation, and the entropy perturbation is dominated by the fluctuation of the bounce field. However, one can see from Eq. \ref{zeta} that by the end of the matter contraction phase, the contribution to the curvature perturbation from each field will become equally important. With this in mind, we follow the evolution of $v_\phi$ and $v_\psi$ during the matter contraction in order to determine the resulting spectrum of $v_\zeta$ at $t_E$. We will see that the result of this is that $\zeta$ acquires a scale invariant spectrum from the bounce field (which was initially the entropy perturbation). This is an explicit realization of the Matter Bounce Curvaton scenario proposed in \cite{Yifu}.

The field fluctuations evolve following the general equations of motion provided in \eqref{pertQ} as analyzed in Appendix A. The perturbation equations can be written in terms of canonical variables
\begin{eqnarray}
 v_{\phi} = a Q_\phi~, ~~v_{\psi} = a Q_{\psi}~ .
\end{eqnarray}
The equations of motion can then be written in Fourier space as
\begin{eqnarray}
v'' _{\phi} + (k^2 - \frac{a''}{a}) v_{\phi} =  J_{\phi \psi} v_{\psi} + J_{\phi \phi} v_{\phi} ~,\\
v'' _{\psi} + (k^2 +m^2 a^2 - \frac{a''}{a}) v_\psi=  J_{\phi \psi} v_{\phi} +  J_{\psi \psi} v_{\psi} ~,
\end{eqnarray}
where the prime denotes the derivative with respect to conformal time, and we define the source (interaction) terms:
\begin{align}
J_{\phi\phi} &=-\frac{9}{2} \Hu^{2} _E\Big(\frac{a_E}{a}\Big)^4 ~,\\
J_{\phi\psi} &= \frac{3}{2} m \Hu_E \frac{a_E}{a}\cos{[ma(\tau-\tilde{\tau}_E)]} ~,\\
J_{\psi\psi} &=\frac{9}{2} \Hu_E^2 \frac{a_E}{a} ~.
\end{align}

We can treat this system perturbatively, using the first order Born approximation to estimate the effect of the source terms. We begin by analyzing the source-free (`homogeneous') system:
\begin{eqnarray}
\label{eom_v_Qphi}
v  _{\phi} ^{(0)}  {''}   + (k^2 -  \frac{a''}{a}) v ^{(0)} _{\phi}   = 0 ~,\\
\label{eom_v_Qpsi}
v _{\psi}  ^{(0)} {''} + (k^2 + a^2 m^2 - \frac{a''}{a}) v ^{(0)} _{\psi}  =0 ~.
\end{eqnarray}
One can see from Eq. \eqref{eom_v_Qphi} that the $k^2$ term will initially dominate, and so the squeezing factor $a^{''}/a$  can be neglected. Thus the dynamics for $v_{\phi}$ corresponds to a free scalar propagating in a flat space-time, and the initial conditions take the form of the Bunch-Davies vacuum:
\begin{eqnarray}
 v_{\phi}^{ini}(\tau, k) \simeq \frac{e^{-ik\tau}}{\sqrt{2k}}~.
\end{eqnarray}
However, the situation for $v_{\psi}$ is different, due to the presence of a non-zero mass. Specifically, in Eq. \eqref{eom_v_Qpsi} when we neglect the last term $a''/a$, the mass term becomes important in addition to the $k^2$ term at the initial moment. Thus one can introduce an effective frequency
for $v_{\psi}$ as
\begin{eqnarray}
 \omega_k^2 = k^2+a^2m^2~,
\end{eqnarray}
and Eq. \eqref{eom_v_Qpsi} has an asymptotic solution which oscillates rapidly with this time dependent frequency on sub-Hubble scales. This is what is expected since the adiabaticity condition $|\omega_k'/\omega_k^2|\ll1$ is satisfied which corresponds to a situation in which the effective physical wavelength is much smaller than the Hubble radius. Therefore, the modes can be regarded as adiabatic when they are in the sub-Hubble regime with $|\omega_k\tau| \gg1$, and we can impose suitable vacuum initial conditions by virtue of a Wentzel-Kramers-Brillouin (WKB) approximation
\begin{equation}
 \sqrt{2\epsilon} ~v_{\psi}^{ini}(\tau, k) \simeq  \frac{1}{\sqrt{2 \omega_k}}  e^{- i \int^\tau \omega_k(\tilde\tau) \mathrm{d}\tilde\tau} ~,
\end{equation}
where $\epsilon\equiv-\dot{H}/H^2=3/2$ in the phase of matter contraction.

During both the matter and Ekpyrotic phases of contraction, the fluctuations modes on scales of cosmological interest today exit the Hubble radius and become classical perturbations \footnote{The classicalization is a consequence of squeezing and decoherence via nonlinear interactions, as discussed in \cite{Martineau, Kiefer}.}.  For matter dominated contraction, one has
\begin{eqnarray}
 a\propto (\tau-\tilde\tau_E)^2~,~~\tilde\tau_E = \tau_E -\frac{2}{{\cal H}_E}~,
\end{eqnarray}
where ${\cal H}_E$ is the conformal Hubble parameter at the moment $t_E$. The gravitational term $a''/a$ leads to the squeezing of field fluctuations. Making use of the vacuum initial condition, we obtain an exact solution to \eqref{eom_v_Qphi}:
\begin{eqnarray}
 {v} _{\phi} ^{(0)} (\tau, k) \simeq \frac{e^{-ik(\tau-\tilde\tau_E)}}{\sqrt{2k}} \left[ 1-\frac{i}{k(\tau-\tilde\tau_E)} \right]~,
\end{eqnarray}
in the phase of matter contraction. For the $v_{\psi}$ mode, there exists a mass term in the expression for the dispersion relation, and thus the field fluctuations do not get squeezed on super-Hubble scales. Instead, one can neglect the $k^2$ term and derive an asymptotical solution as follows,
\begin{eqnarray}
{ v_{\psi}  }  ^{(0)} (\tau, k) \simeq \frac{e^{-iam(\tau-\tilde\tau_E)}}{\sqrt{6am}}~.
\label{EOMpsi}
\end{eqnarray}
These homogenous solutions correspond to a scale invariant spectrum of the entropy mode $\phi$, and a spectrum of the initial curvature mode $\psi$ that is deeply blue:
\begin{eqnarray}
 &&P^{(0)}_\phi\equiv\frac{k^3}{2\pi^2}|\frac{v^{(0)}_{\phi}}{a}|^2 = \frac{H^2}{16\pi^2}~,\\
 &&P^{(0)}_\psi\equiv\frac{k^3}{2\pi^2}|\frac{v^{(0)}_{\psi}}{a\sqrt{3}}|^2 = \frac{k^3}{12\pi^2 m a^3}~.
\end{eqnarray}

As we now show, the entropy mode sources a growing contribution to the curvature mode which then inherits the scale-invariant spectrum of the entropy mode. To compute this effect, we use the 1st order Born approximation in which we evaluate the form of the source terms using the zero'th order solutions. This means that the $1^{st}$ order corrections are determined using
the equation of motion with the following background-dependent source terms:
\begin{eqnarray}
v _{\phi} ^{(1)}  {''}   + (k^2 -  \frac{a''}{a}) v ^{(1)} _{\phi}  &=&  J_{\phi \psi} v^{(0)} _{\psi} + J_{\phi\phi} v^{(0)} _{\phi} ,\\
v _{\psi}  ^{(1)} {''} + (k^2 + a^2 m^2 - \frac{a''}{a}) v ^{(1)} _{\psi}  &=&  J_{\phi \psi} v^{(0)} _{\phi} +  J_{\psi \psi} v^{(0)} _{\psi} ~.
\end{eqnarray}
We solve these for modes on super-Hubble scale and obtain the homogeneous solution plus first order correction,
\begin{eqnarray}
v_\phi &\simeq& v _{\phi} ^{(0)} \left[ 1+ \frac{1}{3} \left ( \frac{a_E}{a}\right)^{3} \right]  +  \frac{ma_E}{3\Hu_E} \left[ 1+ \text{Log} | \frac{2 k}{\mathcal{H}_E}| \right] v _{\psi} ^{(0)} ~,\\
v_\psi &\simeq& v _{\psi} ^{(0)} \left[ 1+ \frac{9\Hu_E}{4 a_E m} \left ( \frac{a_E}{a}\right)^{\frac{1}{2}} \right]  + \frac{3}{2} e^{iam(\tau-\tilde{\tau}_E)} \left ( \frac{a_E}{a}\right)^{\frac{1}{2}} v _{\phi} ^{(0)} ~.
\end{eqnarray}
Correspondingly, the power spectra for two field fluctuations near the end of matter contraction are given by \footnote{Note that the precise form of the mode functions actually includes an arbitrary phase, each of which is drawn from an independent gaussian distribution.  The result of this is that the cross term of $\phi$ and $\psi$ vanishes vanishes when averaged over both distributions to compute the power spectrum. }
\begin{eqnarray}
P_\phi &\simeq&  \frac{16}{9} P_\phi ^{(0)} + \frac{1}{9} \left( \frac{m}{H_E}\right)^2  \left[ 1+ \text{Log} | \frac{2 k}{\mathcal{H}_E} | \right] ^2 P_{\psi}^{(0)}  ~, \nonumber \\
P_\psi &\simeq&  \left[ 1 + \frac{9}{4} \frac{H_E}{m}  \right]^2 P_\psi ^{(0)} + \frac{9}{4} P_{\phi}^{(0)} ~.
\end{eqnarray}
%
We can see from the above expression that the gravitational interaction mixes the spectra of the two fields, such that both fields have a scale invariant piece which is the one which dominates in the infrared.

\subsection{Perturbations in the phase of Ekpyrotic contraction}

During the matter contraction, the energy density of the $\phi$ field becomes more and more important since it is fast rolling along its tachyonic potential. At some moment $t_E$, its contribution to the background energy density starts to dominate over that of the $\psi$ field. We still have $|\phi|\gg1$ and $\dot\phi\ll M_p$ and thus the Lagrangian of $\phi$ is of canonical form with an Ekpyrotic potential. This model then yields an attractor solution of Ekpyrotic contraction
\begin{eqnarray}
 a \propto (\tilde\tau_{B-}-\tau)^{\frac{q}{1-q}}~,~~\tilde\tau_{B-} = \tau_{B-}-\frac{q}{(1-q){\cal H}_{B-}}~.
\end{eqnarray}
We have introduced the instant of time $\tilde\tau_{B-}$ when the scale factor would meet the big crunch singularity if there was no non-singular bounce. If we were not interested in the bounce phase, it would make sense to normalize the time axis such that $\tilde\tau_{B-}=0$, and in this case we would find that the function $g$ would become unity slightly earlier, namely at a time $\frac{q}{(1-q){\cal H}_{B-}}$ (keeping in mind that ${\cal H}_{B-}$ is negative). This signals the beginning moment of the bounce phase $\tau_{B-}$.

Note that, when the universe has not yet arrived at the non-singular bounce phase, the Lagrangian has canonical form and thus the analysis based on gauge invariant field fluctuations (shown in the previous subsection) is still valid. However, one can see that the main contribution to the curvature perturbation has changed from $\delta\psi$ to $\delta\phi$. To render the analysis of cosmological perturbations through the non-singular bounce easier, we switch to the uniform $\phi$ gauge in the Ekpyrotic phase. The detailed analysis of the second order action for perturbations is performed in Appendix B. The simplified quadratic action in this phase is given by:
\begin{align}\label{S_2_ekpy}
 S_{2}=\int d\tau dk^3 \frac{1}{2} \sum_i \bigg[
 v_i'^2 -\Big(k^2 - \frac{q(2q-1)}{(1-q)^2(\tau-\tilde\tau_{B-})^2}\Big)v_i^2 \bigg],
\end{align}
the subscript `$i$' runs over $\{\zeta,\psi\}$. In Appendix B we introduce two new perturbation variables $\{v_\sigma, v_s\}$ which are linear combinations of $v_{\zeta}$ and $v_{\psi}$. This rotation decouples the kinetic terms of $v_{\zeta}$ and $v_{\psi}$ in the general evolution. However, in the model under consider, we can find quadratic actions for $v_\zeta$ and $v_\phi$ which allows for an easy analysis without resorting to a field rotation.

The quadratic action \eqref{S_2_ekpy} yields the following equations of motion for perturbation variables
\begin{eqnarray}\label{EOMcont}
 v_i''+\Big(k^2-\frac{q(2q-1)}{(1-q)^2(\tau-\tilde\tau_{B-})^2}\Big)v_i = 0 ~,
\end{eqnarray}
One can solve for the general solutions to the above equations of motion as follows,
\begin{eqnarray}\label{solution_ekpy}
 v^c_i = C_{i,1} \sqrt{\tau-\tilde\tau_{B-}} J_{\nu_c}(k(\tau-\tilde\tau_{B-})) + C_{i,2} \sqrt{\tau-\tilde\tau_{B-}} Y_{\nu_c}(k(\tau-\tilde\tau_{B-})) ~, \;\;\;\;\;\; i=\zeta,~\psi
\end{eqnarray}
where $\nu_c=\frac{(1-3q)}{2(1-q)}$ and the subscript ``$c$" denotes the Ekpyrotic contracting phase. In addition, $J_{v_c}$ and $Y_{v_c}$ are the two linearly independent Bessel functions with indices $\nu_c$. The coefficients $C_{i,1}$ and $C_{i,2}$ are functions of comoving wave number $k$, and are determined by matching the perturbations at the surface of $t_E$, as we will address in Section \ref{sec:matchingconditions}. For the moment we keep the coefficients general.

Recall that the expression of curvature perturbation $\zeta$ is given by Eq. \eqref{zeta}. When the universe evolves into the Ekpyrotic phase, the trajectory of the background evolution becomes dominated by the bounce field and thus the curvature perturbation is mainly contributed by $Q_\phi$, or equivalently $v_\phi$. Since the matter field $\psi$ no longer dominates over in the background evolution, its field fluctuation $Q_\psi$ plays the role of entropy perturbation.

\subsection{Perturbations through the bounce}

When the bounce field $\phi$ evolves into the range of the ghost condensation, the kinetic term in its Lagrangian is no longer approximately canonical. This triggers a violation of the Null Energy Condition. This causes the universe to exit from the Ekpyrotic phase at some moment $t_{B-}$ and to enter the bounce phase. In this period the bounce field yields a negative contribution to the energy density which will eventually cancel all the other positive contributions, including that of the matter field $\psi$, at a time we denote by $t_B$. We normalize the time axis of the background evolution such that $t_B=0$.
At this moment, the Hubble parameter transits from negative to positive values, crossing $H=0$. As a result, a non-singular bounce takes place.

During the bounce phase, it is a good approximation to model the evolution of the Hubble parameter near the bounce as a linear function of cosmic time:
\begin{equation}
 H(t) = \Upsilon t \, ,
\end{equation}
where $\Upsilon$ is a constant. Such a parametrization is applicable to a wide class of fast bounce models, and the value of $\Upsilon$ depends on the detailed microphysics of the bounce as shown in \eqref{Hubble_bounce}. In addition, the evolution of $\dot\phi$ during the bounce is given by \eqref{dotphi_bouncing}. Making use of the parameterizations for $\dot\phi$ and the Hubble parameter $H$, we can keep the dominant terms of the quadratic action which then simplifies to
\begin{align}\label{ActionBounce}
 S_{2}=\int d\tau dk^3 & \frac{1}{2} \bigg[ v_\zeta'^2- \Big(c^2_\zeta k^2 -\frac{z''}{z}\Big)v_\zeta^2  + v_\psi'^2 - \Big(c^2_{\psi}k^2+a^2 m^2-\frac{a''}{a}\Big)v_{\psi} ^2 \bigg] ~,
\end{align}
where we discuss the role of each term below.

First, we study the gradient terms of the two perturbation modes. The stability of the gradient terms is characterized by the sound speed square parameters, $c_\psi^2$ and $c_\zeta^2$, which are defined in \eqref{soundSpeeds}. In our explicit model, the matter field $\psi$ takes canonical form and thus simply leads to $c_\psi^2 = 1$. Moreover, if we make use of the parameter choice \eqref{parameters1} used in the numerical estimates in the previous section and insert the value of $\dot\phi_B^2$ from \eqref{dot_phi_B^2} as well as the parametrization of the Hubble rate \eqref{Hubble_bounce} into the definition of $c_\zeta^2$, then it takes the following approximate form:
\begin{eqnarray}
 c_\zeta^2 \simeq \frac{1}{3}-\frac{2}{3\sqrt{1+\frac{12\beta V_0}{M_p^4(g_0-1)^2}}} ~,
\end{eqnarray}
in the bounce phase. If we make use of the parameter choice \eqref{parameters1}, we immediately get $c_\zeta^2\simeq-1/3$ which implies that the perturbation $\zeta$ suffers from an gradient instability during the bounce. However, as the duration of the bounce is extremely short, such a exponential growth does not spoil the perturbative control of the analyses \footnote{It does in the bouncing model discussed in \cite{Stein} in which possess a long bounce phase.}.

We have also introduced two quantities to characterize the effective squeezing rates of the perturbation variables
\begin{eqnarray}\label{frequency}
 \frac{a''}{a} \simeq a_B^2(\Upsilon+2\Upsilon^2 t^2) ~, ~~~~
 \frac{z''}{z} \simeq a_B^2 \left[\Upsilon+\frac{2}{T^2} + \Big(2\Upsilon^2+\frac{6\Upsilon}{T^2}+\frac{4}{T^4} \Big)t^2 \right] ~.
\end{eqnarray}
The coefficient $T$ is approximately one quarter of the duration of the bounce phase, and was initially introduced in Eq. \eqref{dotphi_bouncing} to better understand the dynamics of $\dot\phi$ during the bounce. In the limit of a slow bounce, one finds that both squeezing rates are equal which implies that there is no differential growth of the curvature fluctuations relative to the entropy mode across the bounce. In contrast, if we consider a fast bounce model, the gravitational terms $a''/a$ and $z''/z$ differ and lead to enhanced growth of $v_\zeta$ relative to $v_\psi$. However, the overall growth during the bounce phase is bounded from above since the duration of a fast bounce cannot be smaller than  the Planck time if the effective field theory description is to be self-consistent. The bottom line is that given the validity of the effective field theory analysis we can obtain a controllable amplification effect of cosmological perturbations when they evolve through the bounce phase.

The equations of motion for cosmological perturbations during the
bounce phase are given by,
\begin{align}
 v''_\psi +\Big(a^2 m^2+k^2-\frac{a''}{a}\Big)v_\psi = 0 ~,~~~~ v''_\zeta +\Big(c_\zeta^2k^2-\frac{z''}{z}\Big)v_\zeta = 0 ~.
\end{align}
The general solutions to these equations of motion are given by
\begin{eqnarray}
 v^b_\psi(k, \tau)  &=& D_{\psi,1}(k) e^{-\int_{B-} \omega_{\psi}d\tau} +D_{\psi,2}(k) e^{\int_{B-}\omega_{\psi}d\tau} ~, \\
 v^b_\zeta(k, \tau) &=& D_{\zeta,1}(k) e^{-\int_{B-} \omega_\zeta d\tau} +D_{\zeta,2}(k) e^{\int_{B-} \omega_\zeta d\tau} ~,
\end{eqnarray}
with the frequencies $\omega_{\psi}$ and $\omega_{\zeta}$ being
\begin{eqnarray}
\label{omega_psi^2} \omega_{\psi}^2   &\simeq& -k^2 -a_B^2m^2 +a_B^2(\Upsilon+2\Upsilon^2 t^2) ~, \\
\label{omega_zeta^2} \omega_{\zeta}^2 &\simeq& -c_\zeta^2k^2 +a_B^2 \left[\Upsilon+\frac{2}{T^2}+ \Big(2\Upsilon^2+\frac{6\Upsilon}{T^2}+\frac{4}{T^4} \Big)t^2 \right]~,
\end{eqnarray}
respectively. The subscript ``$b$" indicates that we are discussing the solutions in the bounce phase.

Note that we are mainly interested in the infrared modes of cosmological perturbations which are expected to be responsible for the large scale structure of the universe at late times. Therefore, we neglect the $k^2$ terms in the expression for the frequencies and then easily find that $v_\psi$ and $v_\zeta$ are amplified during the bounce phase. Specifically, the amplification factor ${\cal F}_\psi$ for the entropy perturbation $v_\psi$ takes the form:
\begin{eqnarray}\label{F_factor_psi}
 {\cal F}_\psi \equiv e^{\int_{B-}^{B+} \omega_\psi d\tau} \simeq \exp \left [ \Upsilon^{\frac{1}{2}}t + \frac{1}{3}\Upsilon^{\frac{3}{2}}t^3 \right ] \bigg|^{B+}_{B-}~,
\end{eqnarray}
where $B+$ and $B-$ stand for the end and beginning of the bounce phase, respectively. A reasonable bounce model requires $\Upsilon$ to be a very small quantity (which is equivalent to taking the `fast bounce' limit), so that the amplitude of perturbations is in agreement with observations. In this case, the amplification of the entropy mode is in general very small. As a consequence, it is safe to approximately take ${\cal F}_\psi \simeq 1$.

On the other hand, the curvature perturbation experiences an exponential growth through the bounce phase, which can be described by the amplification factor
\begin{eqnarray}\label{F_factor}
 {\cal F}_\zeta \equiv e^{\int_{B-}^{B+} \omega_\zeta d\tau} \simeq \exp \left [ \sqrt{2 +\Upsilon {T^2}}\frac{t}{T} +\frac{2 +3\Upsilon T^2 +\Upsilon^2 T^4}{3 \sqrt{2 +\Upsilon {T^2}}} \frac{t^3}{T^3} \right ] \bigg|^{B+}_{B-} ~.
\end{eqnarray}
This result is exactly the same as the growth factor obtained in the model of single field bounce \cite{BCE}, and thus shows that the amplification effect brought by the effective tachyonic mass term during the bounce is generic.  In the limit of a fast bounce scenario, this amplification factor can be as large as of order $O(10^5)$ as shown in \cite{BCE}. This effect is very important to non-singular bounce cosmologies since such a controllable growth suppresses the tensor-to-scalar ratio, which was originally found to be too large in matter bounce models \cite{Cai:2008qw}.

\subsection{Perturbations in Fast Roll Expansion}

After the bounce, the potential for $\phi$ tends to zero very rapidly. Since the energy density in $\phi$ dominates over the density in $\psi$, this causes us to enter a phase of fast roll expansion, where the quadratic action is given by
\begin{equation}
 S_2 = \int \mathrm{d}\tau \; \mathrm{d}^3 k \frac{1}{2} \sum_i \left[ {v' _i}^2   -(\frac{1}{4 (\tau-\tilde\tau_{B+})^2} + k^2 ) v_i^2  \right]~,
\end{equation}
where the subscript $``i"$ denotes $\zeta$ and $\psi$, respectively. This gives the equations of motion
\begin{equation}
 v_{i}'' + ( k^2 + \frac{1}{4 (\tau-\tilde\tau_{B+})^2} ) v_{i} =0~,
\end{equation}
which yield the solutions
\begin{equation}\label{solution_fastroll}
 v^e_{i} = E_{i,1}(k) \sqrt{\tau-\tilde\tau_{B+}} J_0 \left(k (\tau - \tilde{\tau}_{B+}) \right) + E_{i,2}(k) \sqrt{\tau-\tilde\tau_{B+}} Y_0 \left( k (\tau - \tilde{\tau}_{B+}) \right)~,
\end{equation}
with
\begin{equation}
 a\propto (\tau-\tilde\tau_{B+})^{\frac{1}{2}}~,~~\tilde\tau_{B+}\equiv\tau_{B+}-\frac{1}{2{\cal H}_{B+}}~.
\end{equation}
The subscript ``$e$" indicates that we are discussing the solutions in the fast-roll expanding phase. The coefficients $E_{i,1}(k)$ and $E_{i,2}(k)$ can be determined by matching the perturbations at the moment $\tau_{B+}$. Modulo the square root term, the first mode is constant on super-Hubble scales but the second is growing as a logarithmic function of conformal time. As a consequence, one can see the second term $Y_0$ finally dominates and form the power spectra of cosmological perturbations at late times.

\section{Power spectra of cosmological perturbations}
\label{sec:matchingconditions}

Having solved equations of motion for cosmological perturbations phase by phase, now we are able to study how the solutions can be transferred from initial states to the final ones. We leave the detailed matching processes to Appendix  C and here merely provide a rough description of the analysis.

Our first matching surface is chosen at the moment $\tau_{E}$ where the Ekpyrotic contraction starts and thus is defined by $\rho_\psi=\rho_\phi$. The matching conditions simply require
\begin{equation}
 v^m_{\zeta,\psi}(\tau_{E}) = v_c^{\zeta,\psi}(\tau_{E}) \,\,\, {\rm{and}} \,\,\,
 \frac{d}{d\tau}v^{\zeta,\psi}_{m} (\tau_{E}) = \frac{d}{d\tau}v^{\zeta,\psi}_{m}(\tau_{E}) \, .
\end{equation}
In the Ekpyrotic phase, the growing modes are characterized by the coefficients $C_{\zeta,2}$ and $C_{\psi,2}$ as shown in \eqref{solution_ekpy}, and we focus on super-Hubble scales as it is the long wavelength fluctuations that we are interested in. As a consequence, we can obtain the dominant modes of cosmological perturbations during the Ekpyrotic phase.

Similarly, we match the perturbation modes in the Ekpyrotic contracting phase with those in bounce phase at the moment $\tau_{B-}$. Then we can solve for the coefficients of the growing modes in the bounce phase which are characterized by the coefficients $D_{\zeta,2}$ and $D_{\psi,2}$, respectively. The last matching surface is chosen at the moment $\tau_{B+}$ where primordial cosmological perturbations just pass through the bounce phase and enter the fast-roll expansion. In this case, we are able to determine the forms of $E_{\zeta,2}$ and $E_{\psi,2}$ which are the coefficients of the dominant modes after the bounce.

Substituting the coefficients $E_{\zeta,2}$ and $E_{\psi,2}$ back into the solutions \eqref{solution_fastroll}, we can solve for the asymptotic solutions of the cosmological perturbations in the final stage. On super Hubble scales, these become
\begin{align}
 v^e_\psi &\simeq \frac{\mathcal{ F}_\psi H_E}{2 }\gamma_\psi  e^{-2m/H_E}\Big[ U^{(0)}_\psi\frac{1}{\sqrt{6 a_E m}} +U^{(k)}_\psi\frac{a_E m}{k^{3/2}}\Big] \frac{a_{B-}}{a_{B+}}a(t) ~,\\
 v^e_\zeta &\simeq \frac{\mathcal{ F}_\zeta H_E}{2 }\gamma_\zeta \Big[ U^{(0)}_\zeta\frac{1}{\sqrt{6 a_E m}}+U^{(log)}_\zeta\frac{\text{Log}(\frac{-2k}{a_E H_E})}{\sqrt{6a_E m}} +U^{(k)}_\zeta\frac{a_E m}{k^{3/2}}\Big] \frac{a_{B-}}{a_{B+}}a(t) ~,
\end{align}
where we have defined,
\begin{eqnarray}
 \gamma_\zeta &=& \frac{1}{2(1-3q)} \left[ 1 + \bigg(1 - \frac{\sqrt{2}}{H_{B+} T} (1 + \frac{t_{B+}^2}{T^2} ) \bigg) \ln\frac{a_{B+}}{a(t)}\right ]~, \\
 \gamma_\psi   &=& \frac{1}{2(1-3q)} \left[ 1 + \bigg[1 - \frac{\sqrt{\Upsilon}}{H_{B+} } (1 + \Upsilon{t_{B+}^2} ) \bigg] \ln\frac{a_{B+}}{a(t)} \right]~.
\end{eqnarray}
and the $U$'s are dimensionless coefficient whose detailed form are given in Appendix C.

As a result, we can easily calculate the primordial power spectra of curvature perturbations in the fast roll phase. Up to leading order in $k$, the result is scale invariant,
\begin{align}
 P_\zeta(k) \simeq \frac{k^3}{2\pi^2} \Big| \frac{v^e_{\zeta}}{a}\Big|^2
 \simeq  \frac{\mathcal{ F}_\zeta^2 H_E^2a_E^2}{8\pi^2 }  \gamma_\zeta^2\frac{a_{B-}^2}{a_{B+}^2}(m |U^{(k)}_\zeta|)^2
 \Big[ 1+\mathcal{O}(k^{3/2})\Big] ~.
\end{align}

From the above expression, we can see that the curvature perturbation is dominated by a scale invariant component while there are other terms which can lead to a scale dependence at small length scales. In our model the maximal value of $H_E$ is of the order of the mass parameter $m$, and thus for the perturbation modes which exit the Hubble radius during matter contracting phase the primordial power spectrum is nearly scale-invariant. However, if we consider the perturbation modes on small length scales, the spectrum becomes blue which may lead to interesting observational signals for experiments. 
The absence of a red tilt on large scales indicates that the mechanism for a bounce studied here is 
not the full story, and other ingredients are necessary to have a complete description of cosmology.  We discuss this issue in more detail in the discussion.

\begin{figure}
\includegraphics[scale=0.6]{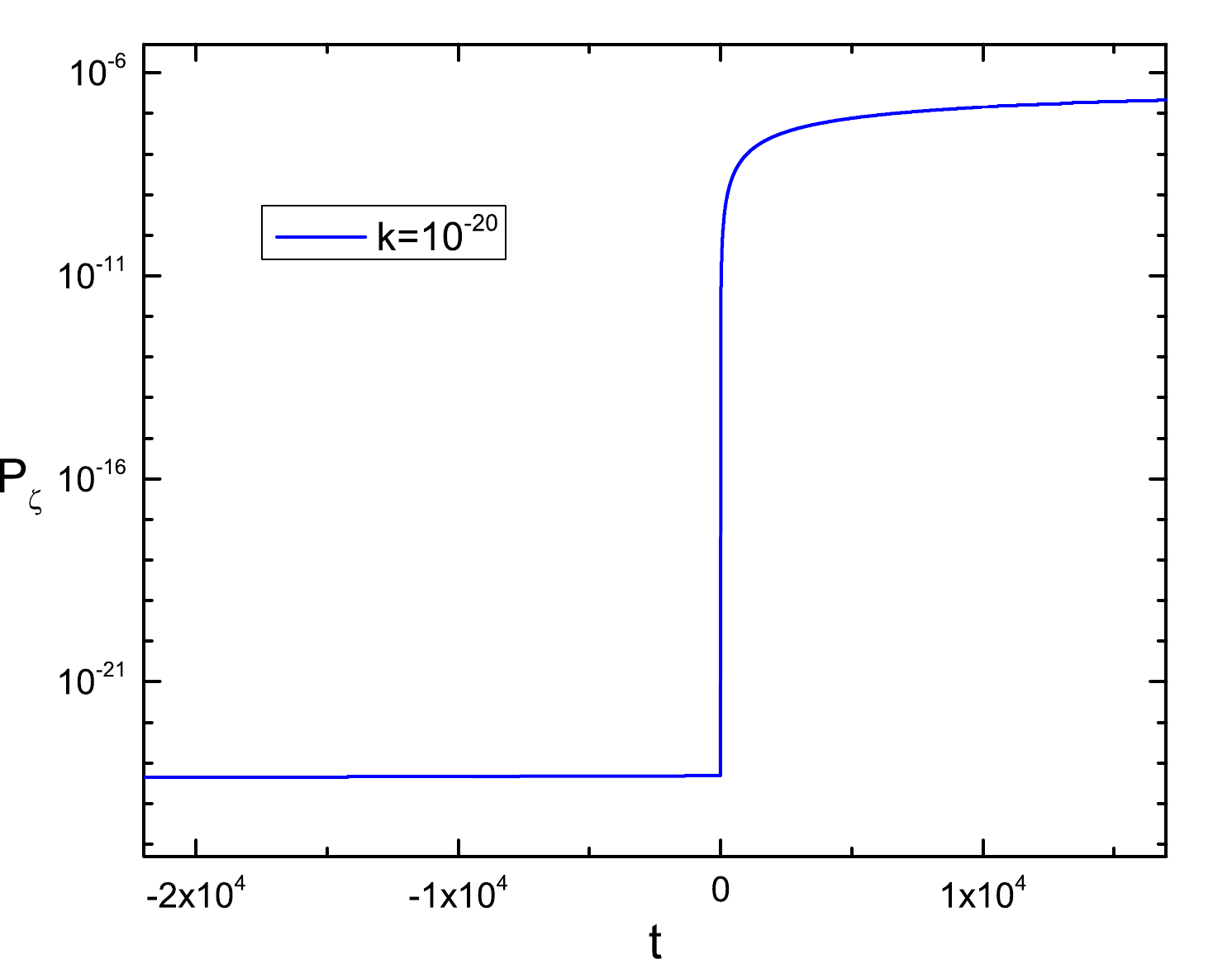}
\caption{Evolution of the power spectrum of curvature perturbation $P_\zeta$ at super-Hubble scale (with a fixed comoving wave number $k=10^{-20}$) as a function of cosmic time. The background parameters are the same as in Fig. \ref{Fig:HEoS}, and the initial condition for the perturbation is chosen as vacuum fluctuation. }
\label{Fig:Pz}
\end{figure}

To provide a check of our analytic calculation of the power spectrum of curvature perturbations, we numerically track its amplitude on super-Hubble scales through the bounce.   From the analytical calculation, we expect the amplitude of curvature perturbation to be conserved before the bounce and to undergo an amplification during the bouncing phase. Specifically, we take the same model parameters as in the background numerics introduced in Section III, and numerically compute the curvature perturbation for a fixed comoving wave number. We show the result in Fig. \ref{Fig:Pz}, in which one can see that the amplitude of curvature perturbations is nearly constant during the contracting phases. During the bounce, the curvature perturbation obtains a dramatic amplification of order $O(10^10)$, corresponding to an amplification factor ${\cal F}_\zeta$ of order $O(10^5)$, in exact agreement with the analytical analysis performed in previous subsections.

The power spectrum of the entropy modes, which is carried (except in the initial matter phase of contraction) by the matter field $\psi$, is also scale-invariant on large scales. It inherits this spectrum from the $\phi$ mode during the matter phase of contraction. However, the amplitude of the entropy mode is negligible in the case of a fast bounce, as the adiabatic mode undergoes a much larger amplification during the bounce.

Thus, this result shows that the universe after the bounce is isotropic and homogeneous and has a nearly scale-invariant spectrum which is almost purely adiabatic. In this sense, our model provides a alternative to inflationary cosmology for explaining the observed spectrum of cosmological perturbations.

We expect the perturbation modes forming the above power spectrum of curvature perturbations will eventually be responsible for the CMB anisotropies. It is therefore interesting to check that the modes relevant to the CMB will exit the Hubble radius during the matter contraction phase. Requiring that the modes do exit the Hubble radius imposes a condition on the bounce. We now perform an estimate of this condition. First, we can write down the wavelengths of the modes exiting the Hubble radius at the beginning and the end of the Ekpyrotic phase associated with today's wavelength $\lambda_0$ as follows,
\begin{align}
 & \lambda(t_{E}) = \frac{a(t_{E})}{a(t_0)}\lambda_0 \simeq \frac{a(t_{E})}{a(t_{B-})} \frac{a(t_{B+})}{a(t_F)} \frac{a(t_{F})}{a(t_0)}\lambda_0~, \\
 & \lambda(t_{B-}) = \frac{a(t_{B-})}{a(t_0)}\lambda_0 \simeq \frac{a(t_{B+})}{a(t_F)} \frac{a(t_{F})}{a(t_0)}\lambda_0~,
\end{align}
respectively, where $t_F$ denotes the end of the Fast-roll expansion with $\rho_\psi(t_F)=\rho_\phi(t_F)$. Recall that the background energy density scales as $\rho \sim a^{-3(1+w)}$, which depends on the background equation of state $w$. Making use of this relation, we derive:
\begin{eqnarray}
 \lambda_{E} \simeq \bigg(\frac{\rho_{B-}}{\rho_E}\bigg)^{\frac{2}{q}} \bigg(\frac{\rho_F}{\rho_{B+}}\bigg)^{\frac{1}{6}} \bigg(\frac{\rho_0}{\rho_F}\bigg)^{\frac{1}{3}} \lambda_0~,\\
 \lambda_{B-} \simeq \bigg(\frac{\rho_F}{\rho_{B+}}\bigg)^{\frac{1}{6}} \bigg(\frac{\rho_0}{\rho_F}\bigg)^{\frac{1}{3}} \lambda_0~.~~
\end{eqnarray}
Since the wavelength of today's observable mode scales as $\lambda_0 \sim t_{eq}$ with $t_{eq}$ being the moment of equality, we require the wavelength of the mode exiting the Hubble radius during the beginning of the Ekpyrotic phase to satisfy
\begin{equation}
 \lambda_E \geq |t_E|~,
\end{equation}
so that the observable modes were generated during matter contraction.

Specifically, we take the density of the universe at present to be $\rho_0 \sim (10^{-12} {\rm GeV})^4$, while at the end of the Fast-roll it is $\rho_F \sim (10^{3} {\rm GeV})^4$, and around the bounce $\rho_B \lesssim (10^{15} {\rm GeV})^4$. From this, we get $\lambda_E \gtrsim 10^{-28}\lambda_0$, which has to be larger than $|t_E|$. This requires $\frac{|t_E|}{t_{eq}} \lesssim 10^{-28}$. Recall that $\rho_m \sim a^{-3} \sim t^{-2}$ during matter contraction, which yields
\begin{eqnarray}
 \frac{\rho_m(t_E)}{\rho_m(t_{eq})} \simeq \bigg( \frac{t_{eq}}{t_E} \bigg)^2 \gtrsim 10^{56}~,
\end{eqnarray}
in the specific case considered above. By inserting the value of the density at matter-radiation equality $\rho_m(t_{eq}) \sim (10^{-9} {\rm GeV})^4$, one can obtain the lower bound on the density of the universe at the end of Ekpyrotic phase,
\begin{eqnarray}
 \rho_m(t_E) \gtrsim (10^{5} {\rm GeV})^4~.
\end{eqnarray}
This condition needs to be satisfied in order for the modes exiting the Hubble radius during matter contraction to be responsible for the CMB anisotropies. Note that if we consider a bounce at lower energy scales then the above condition to be satisfied for a large portion of parameter space .

\section{Tensor perturbation}

Similar to scalar modes,  tensor perturbations are generated from vacuum fluctuations on sub-Hubble scales in the matter-dominated contracting phase. As the universe contracts, the tensor modes exit the Hubble radius. As is well known, the equation of motion for the tensor fluctuations is the same as that of a massless scalar field. Hence, vacuum initial conditions lead to the same amplitude of the tensor modes and the curvature fluctuations on sub-Hubble scales. Once on
super-Hubble scales, the tensor modes are squeezed. During the phases of Ekpyrotic contraction, bounce and fast-roll expansion the equation of motion for the tensor modes is the same as that for the entropy mode (in the absence of mass for the latter). In particular, the squeezing factor of the modes is $a'' / a$. As we showed above, the amplitude of the entropy mode at the beginning of the Ekpyrotic phase is of the same order as that of the curvature modes, which in turn is the same order as that of the tensor modes. After the beginning of the Ekpyrotic phase the tensor and entropy modes evolve the same way. Therefore, it is easy to derive the power spectrum of primordial tensor modes. Making use of the expression \eqref{u_e^h}, one obtains the following expression for the power spectrum of primordial tensor perturbations:
\begin{eqnarray}
 P_T\equiv \frac{k^3}{2\pi^2} |\frac{u_h}{a}|^2 \simeq \frac{{\cal F}_{\psi}^2 \gamma_\psi^2H_E^2}{64\pi^2 M_p^2(2q-3)^2} \frac{a_{B-}^2}{a_{B+}^2} ~.
\end{eqnarray}
One can see the power spectrum of primordial tensor modes in our model is scale-invariant. This spectrum  is inherited from the power spectrum of primordial curvature perturbations on large scales. The evolution and the amplification during the bounce phase, however, follow the behavior of entropy perturbations.

One can define a tensor-to-scalar ratio,
\begin{eqnarray}
 r_T \equiv \frac{P_T}{P_\zeta} \simeq \frac{{\cal F}_{\psi}^2\gamma_{\psi}^2}{{\cal F}_\zeta^2\gamma_{\zeta}^2}~.
\end{eqnarray}
This ratio is given by the ratio of amplification factors of curvature and entropy modes during the bounce phase. Thus, this ratio can be greatly suppressed by a large value of the factor ${\cal F}_\zeta$. Considering the group of canonical values for model parameters as given in the previous section discussing the background analysis, we find that this ratio can be as low as of order $O(10^{-8})$.

\section{Conclusion and Discussion}

We have studied the evolution of the background and of the linear cosmological fluctuations in a two field matter bounce model in which one field ($\psi$) represents the regular matter which has a time-averaged equation of state $p = 0$, and the second field ($\phi$) is responsible for both an Ekpyrotic phase of contraction which follows the initial matter-dominated period, and which yields a non-singular bounce. As a consequence of the Ekpyrotic phase of contraction, there is no BKL instability in this model \footnote{The BKL instability to the growth of anisotropies is a problem which afflicts most bouncing cosmological models.}. Thus, as long as the initial conditions are chosen such that the Ekpyrotic period of contraction begins before the anisotropies dominate, the background will evolve towards a homogeneous and isotropic state.

Since there are two matter fields present, it is important to study not only the adiabatic fluctuations (as was done in \cite{BCE}), but also the entropy mode. We have shown that in the matter phase of contraction the adiabatic mode (which is seeded by the massive field $\psi$) starts out with a deep blue spectrum, and it is only the entropy mode (which is seeded by
the effectively massless field $\phi$) which acquires a scale-invariant spectrum via squeezing on super-Hubble scales during the phase of matter contraction. However, the entropy mode continuously seeds a contribution to the curvature fluctuation. This contribution is scale-invariant, and we have shown that its amplitude at the end of the matter phase of contraction is of the same order of magnitude as the initial entropy fluctuation. Once the Ekpyrotic phase of contraction begins, the roles of the adiabatic and entropy modes change: it is now the dominant field $\phi$ which determines the adiabatic mode, and $\psi$ becomes the entropy mode. Since the fluctuations in $\phi$ have a scale-invariant spectrum, the curvature perturbations inherit a scale-invariant spectrum at the beginning of the Ekpyrotic phase, whereas the fluctuations associated with $\psi$ which have developed a scale-invariant form (due to the seeding mentioned above) become the entropy mode.

We followed the evolution of both the adiabatic and the entropy modes from the beginning of the Ekpyrotic phase of contraction through the non-singular bounce phase and into the following fast-roll phase of expansion. Both modes preserve their scale-invariant spectrum. The curvature fluctuations are amplified during the bounce phase, but for a fast bounce the amplification of the entropy mode is negligible. Hence, the entropic contribution to the late time fluctuations is suppressed. It is, in fact, suppressed by the same factor as the tensor perturbations to the scalar ones, since the tensor modes have the same squeezing factor as the entropy field.

One serious shortcoming of the model under consideration is the lack of a prediction for the spectral tilt of perturbations in agreement with CMB observations, which require a red tilt. At best, this model is capable of producing a scale invariant spectrum for the CMB, however scale invariance has been ruled out by Planck at the $5\sigma$ level \cite{Ade:2013uln}. Given this, we emphasize that the focus of our work is the study of perturbations through a non-singular bounce, and hence we are primarily concerned with the evolution inside of  the `black-box' that separates the contracting and expanding branches of the cosmological evolution. The tilt is due to the choice of contracting branch, and in this study we have chosen matter contraction as our toy model, purely for the sake of simplicity.

However, there do exist mechanisms which could induce a red tilted spectrum in this model. The simplest possibility is to generate the red tilt via a tachyonic coupling to a curvaton field. The effect of curvatons in a matter bounce was originally investigated in Ref. \cite{Yifu}, where one can quickly see that a tachyonic coupling $g^2<0$ will cause the the spectral index in eq. (23) of \cite{Yifu} to be red, without generating any instability. Another mechanism is to change the matter field to a fluid with slightly negative pressure, as was mentioned in \cite{WilsonEwing:2012pu}. We plan to investigate these mechanisms in future work.

Finally, we would like to comment on the reheating process. In this specific model under consideration, we assume the two fields are only coupled through gravitational interactions. Therefore it is straightforward to track the evolution of both the background and perturbation modes. In a more generic case, the universe described by our model can be reheated by several different methods, e.g. the usual treatment of reheating in the fast roll phase, perturbative decay of the bounce field in Ekpyrotic phase, and gravitational particle production during a phase transition such as the bouncing phase \cite{Cai:2011ci}. Another mechanism of reheating the universe it to introduce a kinetic coupling such as was done for the defrosting process in an emergent galileon cosmology \cite{LevasseurPerreault:2011mw}. This aspect, as well as the comparison with the CMB data, provides us with quite a few interesting topics which we will explore in future work.

\begin{acknowledgments}

This work is supported in part by an NSERC Discovery grant and by funds from the Canada Research Chair program.

\end{acknowledgments}

\appendix

\section{Cosmological perturbations in a double field model of canonical form}

In this Appendix we shall review the equations of motion for the coupled curvature and entropy modes in a model with two canonical scalar fields. Particular focus is on the curvature modes induced by an initial mode. We will apply this theory to the matter-dominated phase of contraction during which both of the scalar fields in our model have kinetic terms in the action which are approximately canonical. Note that the matter fields $\phi_i$ considered below have mass dimension one, and hence to apply these formulas our Galileon field $\phi$ must be multiplied by $M_p$.

We shall work in longitudinal gauge in which the linearized scalar metric fluctuations appear in the metric in the following way (see e.g. \cite{MFB, RHBpertrev}):
\begin{eqnarray}
 ds^2 \, = \, (1+2\Phi)dt^2-a^2(t)(1-2\Psi)d\vec{x}^2~,
\end{eqnarray}
where $t$ is cosmic time and $x^i$ are the comoving spatial coordinates. The scalar metric fluctuations
are characterized by two functions $\Phi$ and $\Psi$ which depend both on space and time. We take
matter to consist of a set of scalar fields $\phi_i$, which in our explicit model are the bounce field
$\phi$ and the matter field $\psi$. If the gravitational action is the usual one, then the matter
sector does not admit linearized anisotropic stress the off-diagonal components of the perturbed
Einstein equations imply $\Psi=\Phi$. By expanding the Einstein and matter equations to first order,
we obtain the following perturbation equations:
\begin{eqnarray}
\delta\ddot\phi_i + 3H\delta\dot\phi_i + [-\frac{\nabla^2}{a^2}\delta\phi_i
+ \sum_jV_{,ij}\delta\phi_j]
&=& 4\dot\phi_i\dot\Phi-2V_{,i}\Phi~, \\
-3H\dot\Phi + (\frac{\nabla^2}{a^2}-3H^2)\Phi
&=& 4\pi{G}\sum_i[\dot\phi_i\delta\dot\phi_i-\dot\phi_i^2\Phi+V_{,i}\delta\phi_i]~,\\
\dot\Phi + H\Phi  &=& 4\pi{G}\sum_i\dot\phi_i\delta\phi_i~,
\end{eqnarray}
where $V_{,i}$ denotes the derivative of the scalar field potential with respect to $\phi_i$.

We can recast the above equations in terms of the Sasaki-Mukhanov variables
\cite{Sasaki:1986hm, Mukhanov:1988jd} which are defined as
\begin{eqnarray}
Q_i\equiv\delta\phi_i+\frac{\dot\phi_i}{H}\Phi~,
\end{eqnarray}
and in terms of which the equations of motion are given by \cite{NT, Gordon, FB}
\begin{equation}\label{pertQ}
\ddot{Q}_i +3H\dot{Q}_i -\frac{\nabla^2}{a^2}Q_i
+\sum_j[V_{,ij}-\frac{1}{a^3M_p^2} \frac{\mathrm{d}}{\mathrm{d}t}(\frac{a^3}{H}\dot\phi_i\dot\phi_j)]Q_j = 0~.
\end{equation}

To combine the above equations, one can define the quantity $\zeta$ which is the curvature perturbation
on the uniform density slice,
\begin{eqnarray}
 \zeta = H\frac{\sum_i \dot\phi_i Q_i}{\sum_j \dot\phi_j^2}~ .
\end{eqnarray}
This quantity is conserved on super-Hubble scales in an expanding universe if there are only adiabatic
fluctuations \cite{BST, BK}. However, the presence of entropy fluctuations on large scales will lead to a growth of
$\zeta$ which corresponds to the seeding of an adiabatic fluctuation mode by the entropy mode.

At linear order, the equation for the time derivative of $\zeta$ in the case of two matter fields
$\phi$ and $\psi$ (both with mass dimension one) is given by \cite{NT, Gordon}
\begin{eqnarray}\label{dotzeta}
 \dot\zeta = -\frac{H}{\dot{H}} \frac{\nabla^2}{a^2} \Phi  -
 \frac{H}{2} \bigl( \frac{\delta \phi}{{\dot \phi}} - \frac{\delta \psi}{{\dot \psi}} \bigr)
\frac{\mathrm{d}}{\mathrm{d}t} \bigl( \frac{ {\dot \phi}^2 - {\dot \psi}^2}{ {\dot \phi}^2 + {\dot \psi}^2} \bigr) ~.
\end{eqnarray}
On large scales, the first term of the r.h.s of Eq. (\ref{dotzeta}) is negligible. The second term describes
the transfer of entropy to adiabatic fluctuations, the term we are interested in.

\section{General second order action for cosmological perturbations in uniform $\phi$ gauge}

It is useful to study perturbation theory by making use of the ADM metric. Particularly, we focus on
the part of the action involving the scalar metric perturbation $\zeta$ and the matter field fluctuations
$\delta\phi$ and $\delta\psi$. It is well known that one scalar degree of freedom can be fixed by
a gauge choice. We choose the following uniform field gauge:
\begin{eqnarray}
 \delta\phi=0~,~~h_{ij}=a^2e^{2\zeta}\delta_{ij}~.
\end{eqnarray}

After a lengthy calculation, the Lagrangian \eqref{Lagrangian} expanded to quadratic order in
the fluctuations becomes
\begin{align}\label{secorderaction}
 S_{2}=&\int dt dx^3 a(t)^3 \Big[(2 M_p^2 \dot\zeta - 2M_p^2 H\alpha+\dot\phi^3G_{,X}\alpha+\dot\psi P_{,Y} \delta\psi ) \frac{\partial_i^2\sigma}{M_p^2 a^2}\nonumber\\
 &-3M_p^2\dot\zeta^2-2M_p^2 \alpha\frac{\partial_i^2\zeta}{a^2}+6M_p^2H\alpha\dot\zeta-3\dot\phi^3G_{,X}\alpha\dot\zeta\nonumber\\
 &+M_p^2\frac{(\partial_i\zeta)^2}{a^2}-3M_p^2 H^2\alpha^2+\frac{\dot\phi^2}{2}K_{,X}\alpha^2\nonumber\\
 &+\frac{\dot\phi^4}{2}K_{,X X}\alpha^2+6H\dot\phi^3 G_{,X}\alpha^2+\frac{3}{2}H\dot\phi^5 G_{,X X}\alpha^2\nonumber\\
 &-\dot\phi^2(G_{,\phi}+\frac{\dot\phi^2}{2}G_{,X \phi})\alpha^2 +3\zeta(\delta\psi P_{,\psi}+\dot\psi P_{,Y}\dot{\delta\psi})\nonumber\\
 &+\frac{\dot\psi^2}{2}P_{,Y}\alpha^2+\frac{\dot\psi^4}{2}P_{,Y Y}\alpha^2\nonumber\\
 &+\alpha \left(\delta\psi(P_{,\psi}-\dot\psi^2 P_{,Y \psi})-\dot{\delta\psi}\dot\psi(P_{,Y}+\dot\psi^2 P_{,Y Y}) \right)     \Big]~,
\end{align}
where $\alpha$ and $\partial_i\sigma$ are the lapse function and shift vector, respectively.
Varying the quadratic action \eqref{secorderaction} with respect to $\alpha$ and $\sigma$ yields
\begin{align}\label{alpha}
\alpha=\frac{2M_p^2 \dot\zeta+\dot\psi P_{,Y}\delta\psi}{2M_p^2H-\dot\phi^3 G_{,X}} ~,
\end{align}
as well as the expression for $\sigma$.

Substituting $\alpha$ and $\sigma$ back into the action, we then obtain a much simplified form
\begin{align}\label{secorderaction2}
 S_2=\int dt dx^3 &\left\{ \frac{a}{2} z^2\Big[ \dot\zeta^2-\frac{{c_\zeta}^2}{a^2}(\partial_i\zeta)^2  \Big] \right. \nonumber\\
 &  +\frac{a}{2}y^2\Big[\dot{\delta\psi}^2- \frac{{c_\psi}^2}{a^2}(\partial_i\delta\psi)^2 +\frac{2}{ay^2}M^2_{\delta\psi}\delta\psi^2\Big]    \nonumber \\
 & \left. +C_{1}\delta\psi\zeta +C_{2}\delta\psi\dot\zeta + C_{3}{\delta\dot\psi}\dot\zeta+C_{4}\partial^i\delta\psi\partial_i\zeta  \right\}~,
\end{align}
where $C_{1,2,3,4}$ are the coefficients in front of the interaction terms
\begin{align}
 C_1 &= 3a^3\dot\psi^2 \Big(\ddot\psi P_{,YY}-P_{,Y\psi}\Big)~,\\
 C_2 &= \frac{2M_p^2a^3}{(2M_p^2H-\dot\phi^3 G_{,X})^2} \Big( 12\dot\phi^3\dot\psi H G_{,X}P_{,Y} -6M_p^2\dot\psi H^2P_{,Y} +\dot\phi^2\dot\psi K_{,X}P_{,Y} \nonumber\\
 & +\dot\psi^3{P_{,Y}}^2 +3\dot\phi^5\dot\psi HG_{,XX}P_{,Y} +\dot\phi^4\dot\psi K_{,XX}P_{,Y} +\dot\psi^5P_{,Y}P_{,YY} \nonumber\\
 & -2\dot\phi^2\dot\psi G_{,\phi}P_{,Y} +2M_p^2HP_{,\psi} -\dot\phi^3G_{,X}P_{,\psi} -\dot\phi^4\dot\psi G_{,X\phi}P_{,Y} \nonumber\\
 & -2M_p^2\dot\psi HP_{,Y\psi} +\dot\phi^3\dot\psi^2G_{,X}P_{,Y\psi} \Big)~,\\
 C_3 &= -\frac{2M_p^2a^3\dot\psi}{2M_p^2H-\dot\phi^3 G_{,X}} \Big(P_{,Y}+\dot\psi2P_{,YY}\Big)~,\\
 C_4 &= \frac{2M_p^2a\dot\psi P_{,Y}}{2M_p^2H-\dot\phi^3 G_{,X}} ~.
\end{align}
The parameters $z^2$ and $y^2$ are defined as the coefficients of $\dot{\phi}^2$ and $\dot{\psi}^2$ respectively, and are given by
\begin{align}\label{yz}
 z^2&=\frac{4M_p^4 a^2}{(2M_p^2H-\dot\phi^3 G_{,X})^2}\Big( 6H\dot\phi^2 G_{,X}+\frac{3}{2M_p^2}\dot\phi^2 G_{,X}+\dot\phi^2 K_{,X}+\dot\phi^4 K_{,X X}\nonumber\\
 & +\dot\psi^2 P_{,Y}+\dot\psi^4 P_{,Y Y}+3H\dot\phi^5 G_{,X X}-2\dot\phi^2 G_{,\phi}-\dot\phi^4 G_{,\phi X} \Big)\\
 y^2&=a^2\Big(P_{,Y}+\dot\psi^2 P_{,Y Y} \Big) ~.
\end{align}
The sound speeds of $\zeta$ and $\psi$ are denoted $c_{\psi,\zeta}$, and are given by
\begin{align}\label{soundSpeeds}
{c_{\psi}}^2 &= \frac{P_{,Y}}{P_{,Y}+\dot\psi^2 P_{,Y Y}} ~, \\
{c_{\zeta}}^2 &= \frac{2 a^2}{z^2} \bigg[ M_p^2 - \frac{3M_p^4 H}{2M_p^2H - \dot\phi^3 G_{,X}} - \frac{3M_p^4 ( 3\dot\phi^2\ddot\phi G_{,X}+\dot\phi^4 G_{,\phi X}+\dot\phi^4 \ddot\phi G_{,X X}-2M^2 \dot H ) }{(2M_p^2H-\dot\phi^3 G_{,X})^2}  \bigg] ~.
\end{align}

One can define canonical variables for perturbation modes $\zeta$ and $\psi$ as follows,
\begin{eqnarray}\label{v_zeta,psi}
 v_\zeta \equiv z\zeta~,~~ v_\psi \equiv y\delta\psi~,
\end{eqnarray}
and then the time derivative terms in the quadratic action become of canonical form in conformal coordinates.

\section{Matching Coefficients}\label{matchingappendix}

The matching conditions for cosmological perturbations were discussed in \cite{HV, DM}. The idea was
to match two solutions of General Relavitity across some space-like matching surface which is endowed
with the localized stress-energy to enable the transition between the two space-times. The
matching conditions state that the induced metric on the matching surface must be the same when
calculated from either side, and that the extrinsic curvature jumps by an amount given by the localized
stress-energy on the surface.

In our case, the background is continuous across the various matching surfaces (this would not have
been the case had we cut out the bouncing phase and tried to match directly between the contracting
Ekpyrotic phase and the expanding fast-roll period). Hence, there is no jump in the extrinsic
curvature across the matching surface. Matter fields must also evolve continuously across the
bounce. Hence, the matching conditions are
\begin{equation}
v_\zeta^1(\tau_m) = v_\zeta^2(\tau_m) \,\,\, {\rm{and}} \,\,\,
\frac{d}{d\tau}v_\zeta^1(\tau_m)=\frac{d}{d\tau}v_\zeta^2(\tau_m) ~,
\end{equation}
at each matching time $\tau_m$, where the superscripts $1$ and $2$ indicate the values of the
variables computed in the phases after and before the matching surface, respectively, and
\begin{equation}
v_\psi^1(\tau_m) = v_\psi^2(\tau_m) \,\,\, {\rm{and}} \,\,\,
\frac{d}{d\tau}v_\psi^1(\tau_m)=\frac{d}{d\tau}v_\psi^2(\tau_m) ~.
\end{equation}

We first match the cosmological perturbation $v_\zeta$ and $v_\psi$ at the beginning moment of
the Ekpyrotic phase $\tau_E$.
The matching conditions allow us to determine the dominant coefficients $C_{i,2}$ in the Ekpyrotic phase,
with the result
\begin{eqnarray}
 C_{i,2} &\simeq&\frac{\pi\sqrt{\tau_E-\tilde{\tau_{B-}}}}{2v_c \Gamma_{v_c}}\Big(\frac{k (\tau_E-\tilde{\tau}_{B-})}{2}\Big)^{v_c}\Big[\frac{d}{dt}v^m_i-v^m_i \frac{1+2v_c}{2(\tau_E-\tilde{\tau}_{B-})}  \Big] \, .
\end{eqnarray}

The coefficients $D_{i,1}$ and $D_{i,2}$ for the solutions in the bounce phase are also derived by
matching $v_\zeta$ and $v_\psi$ at the end moment of Ekpyrotic phase $\tau_{B-}$. Picking out
the dominant terms yields
\begin{eqnarray}
 D_{\zeta,2} &\simeq& \frac{ -C_{\zeta,2} \Gamma_{\nu_c} e^{\int^{\tau_{B+}}_{\tau_{B}} w_\zeta d\tau} }{ 2^{2-\nu_c} \pi \omega_\zeta k^{\nu_c} (\tau_{B-}-\tilde\tau_{B-})^{\frac{1}{2}+\nu_c} } \bigg[1-2\nu_c+2\omega_\zeta(\tau_{B-}-\tilde\tau_{B-}) \bigg]
 ~, \\
 D_{\psi,2} &\simeq& \frac{ -C_{\psi,2} \Gamma_{\nu_c}e^{\int^{\tau_{B+}}_{\tau_{B}} w_\psi d\tau} }{ 2^{2-\nu_c} \pi \omega_\psi k^{\nu_c} (\tau_{B-}-\tilde\tau_{B-})^{\frac{1}{2}+\nu_c} } \bigg[ 1-2\nu_c+2\omega_\psi(\tau_{B-}-\tilde\tau_{B-}) \bigg]~.
\end{eqnarray}

After the bounce, we match the cosmological perturbations at the moment $\tau_{B+}$ and then determine the coefficients $E_{\zeta,i}$ and $E_{\psi,i}$. Both are important so we write them all,
\begin{align}
E_{1,i}&=-D_{i,2}\frac{e^{\int^{\tau_{B+}}_{\tau_{B}} w_i d\tau}}{2\sqrt{\tau_{B+}-\tilde{\tau}_{B+}}} \Bigg[-2+\big(2(\tau_{B+}-\tilde{\tau}_{B+})w_i-1\big)\Big(\text{ln}\Big[\frac{k(\tau_{B+}-\tilde{\tau}_{B+})}{2}\Big]+\gamma_E\Big)\Bigg] ~,\\
E_{2,i}&=D_{i,2} \frac{\pi e^{\int^{\tau_{B+}}_{\tau_{B}} w_i d\tau}}{4\sqrt{\tau_{B+}-\tilde{\tau}_{B+}}} \bigg[1-2(\tau_{B+}-\tilde{\tau}_{B+}) \bigg] ~.
\end{align}
By making use of these coefficients, we can extract the dominant mode of cosmological perturbations in
the fast-roll expanding phase, namely
\begin{align}
v^e_\psi &\simeq \frac{\mathcal{ F}_\psi H_E}{2 }\gamma_\psi  e^{-2m/H_E}\Big[ U^{(0)}_\psi\frac{1}{\sqrt{6 a_E m}} +U^{(k)}_\psi\frac{a_E m}{k^{3/2}}\Big] \frac{a_{B-}}{a_{B+}}a(t) ~,\\
v^e_\zeta &\simeq \frac{\mathcal{ F}_\zeta H_E}{2 }\gamma_\zeta \Big[ U^{(0)}_\zeta\frac{1}{\sqrt{6 a_E m}}+U^{(log)}_\zeta\frac{\text{Log}(\frac{-2k}{a_E H_E})}{\sqrt{6a_E m}} +U^{(k)}_\zeta\frac{a_E m}{k^{3/2}}\Big] \frac{a_{B-}}{a_{B+}}a(t) ~,
\end{align}
with
\begin{align}
U^{(k)}_\zeta&=-(25+49q) i\frac{H_E}{24m}-\frac{27}{24}q,~~~U^{(log)}_\zeta=\sqrt{2}\frac{m}{H_E}\Big(1-\frac{5}{2}q\Big)\nonumber\\
U^{(0)}_\zeta&=\sqrt{2}\Big(1-\frac{3}{2}q-\frac{27}{8}iq+\frac{9H_E}{8m}(1-q)+ \frac{m}{3H_E}\big(1-q-9iq\big)\Big)\nonumber\\
U^{(k)}_\psi&=-\frac{3}{8}\Big(\sqrt{3}(1-q)\frac{H_E}{m}-3q\Big),~~~U^{(0)}_\psi=1+\frac{9}{2}\frac{H_E}{m}-\big(\frac{3}{2}-\frac{27}{8}i\big)q-\frac{9}{8}\frac{H_E}{m}q-3i\frac{mq}{H_E}.\nonumber
\end{align}
We have also defined the constants $\gamma_\zeta$ and $\gamma_\psi$ as the coefficients who comes from the asymptotic form of the Bessel function $Y_0$ on large length scales,
\begin{eqnarray}
 \gamma_\zeta &=& \frac{1}{2(1-3q)} \left[ 1 + \bigg(1 - \frac{\sqrt{2}}{H_{B+} T} (1 + \frac{t_{B+}^2}{T^2} ) \bigg) \ln\frac{a_{B+}}{a(t)}\right ]~, \\
 \gamma_\psi   &=& \frac{1}{2(1-3q)} \left[ 1 + \bigg[1 - \frac{\sqrt{\Upsilon}}{H_{B+} } (1 + \Upsilon{t_{B+}^2} ) \bigg] \ln\frac{a_{B+}}{a(t)} \right]~.
\end{eqnarray}

Similarly, one can track the evolution of primordial tensor modes and determine the matching relations.
Comparing with the evolution of entropy perturbation, the tensor fluctuations differ only in the mass term
and the choice of initial conditions which only affects the evolution before the Ekpyrotic phase. During
and after the phase of Ekpyrotic contraction, the evolution of entropy perturbations and tensor fluctuations
are described by the same equation of motion. Working at the level of homogeneous solutions, the tensor fluctuations will have the same amplitude as $v_\phi$ at the end of the matter contraction phase
(both come from a massless field that has the same vacuum amplitude). Hence we conclude that the final
amplitude of the tensor modes will be of the form,
\begin{eqnarray}
\label{u_e^h}
 u^e_h(k,\tau) &\simeq& - \frac{ i {\cal F}_{\psi} \gamma_\psi \mathcal{H}_E } {4 a_E \sqrt{2k^3}(2q-3)} \frac{a_{B-}}{a_{B+}} a(\tau)~,
\end{eqnarray}
in the fast-roll expanding phase.

\end{document}